\documentstyle[eqsecnum,preprint,aps,prd]{revtex}
\tighten
\begin{document}
\preprint{WISC-MILW-94-TH-11}
\draft

\title{CBR Anisotropy from Primordial Gravitational Waves in
Inflationary Cosmologies}

\author{Bruce Allen}
\address{
Department of Physics, University of Wisconsin -- Milwaukee\\
P.O. Box 413, Milwaukee, Wisconsin 53201, U.S.A.\\
email: ballen@dirac.phys.uwm.edu}
\author{Scott Koranda}
\address{
Department of Physics, University of Wisconsin -- Milwaukee\\
P.O. Box 413, Milwaukee, Wisconsin 53201, U.S.A.\\
email: skoranda@dirac.phys.uwm.edu}
\date{}
\maketitle
\begin{abstract}
We examine stochastic temperature fluctuations of the cosmic background
radiation (CBR) arising via the Sachs-Wolfe effect from gravitational
wave perturbations produced in the early universe.  These temperature
fluctuations  are described by an angular correlation function
$C(\gamma)$.  A new (more concise and general) derivation of
$C(\gamma)$ is given, and evaluated for inflationary-universe
cosmologies.  This yields standard results for angles $\gamma$ greater
than a few degrees, but new results for smaller angles, because we do
not make standard long-wavelength approximations to the gravitational
wave mode functions.  The function $C(\gamma)$ may be expanded in a
series of Legendre polynomials; we use numerical methods to compare the
coefficients of the resulting expansion in our exact calculation with
standard (approximate) results.   We also report some progress towards
finding a closed form expression for $C(\gamma)$.
\end{abstract}


\section{INTRODUCTION}

\label{section1}

Penzias and Wilson \cite{PenziasWilson} discovered the Cosmic
Background Radiation (CBR) in 1965.  Since then researchers have
studied the CBR using ground, balloon, rocket, and satellite based
experiments \cite{experiments,SmootEtAl}.  The evidence indicates that
this radiation is a remnant of an early hot phase of the universe,
emitted when  ionized hydrogen and electrons combined at a temperature
of about 4000 K \cite{KolbTurner}.  In the simplest models this
combination occurs at a redshift $Z \approx 1300$, although it is also
possible that the hydrogen was re-ionized as recently as redshift $Z
\approx 100$ \cite{JonesWyse}.  In effect, the CBR is a picture of our
universe when it was much smaller and hotter than it is today.

The CBR has a thermal (blackbody) spectrum, and is remarkably isotropic
and uniform.  Only recently have experiments reliably detected
perturbations away from perfect isotropy.  Such perturbations are
expected; in 1967 Sachs and Wolfe \cite{SachsWolfe} showed how
variations  in the density of the cosmological fluid and gravitational
wave perturbations result in CBR temperature fluctuations, even if the
surface of last scattering was perfectly uniform in temperature.

During the past several years, the Cosmic Background Explorer (COBE)
satellite team has reported detailed measurements of the statistical
properties of these temperature perturbations \cite{SmootEtAl}.
Analyzing the COBE data is subtle; it requires subtraction of the
dipole and quadrupole moments arising from the Doppler shift due to the
earth's peculiar velocity w.r.t. the cosmological fluid, and also the
subtraction of infra-red and microwave emission from stars, dust clouds
and gas within our own galaxy.  In this paper, we assume that these
contaminants have been  removed from the data, and discuss only the
perturbations of the CBR which are cosmological in origin.

Additional measurements by other experimental groups
\cite{MSAM,TEN,SP91,SK,OVRO,MAX,WD2,PYTHON} have also reported
perturbations of the CBR over variety of angular scales.  The range of
angular scales covered by these different experiments is nicely
illustrated in Figure 1b of reference \cite{Critt2}; the angular scales
range from full sky coverage (180 degrees) down to angular scales less
than 1/10 of a degree.  These experiments are ongoing, and additional
data should appear from these research groups over the next few
years.

For our purpose, the most useful statistical quantity determined by
COBE, and the device frequently used to state and compare the results
of the other experiments, is the sky-averaged angular correlation
function \begin{equation} C(\hat u^a,\hat v^b) = C(\gamma)=\langle
{\delta T \over T}(\hat u^a) {\delta T \over T}(\hat v^b)  \rangle_{\rm
sky} .  \label{classcorr} \end{equation} In this formula, $\hat u^a$
and $\hat v^b$ are two unit-length spatial vectors, pointing out from
the observer's location to points on the celestial sphere.   The CBR
temperature fluctuation in the direction $\hat u^a$ away from the mean
value of $T$ is denoted $\delta T(\hat u^a)$.  The angle brackets, as
used in individual experiments, refer to a uniform ``sky average" over
all points on the celestial sphere separated by angle $\gamma$, where
$\cos \gamma = \hat u^a \hat v_a$.  For this reason, the correlation
function depends only on the angle $\gamma$, and not on the absolute
position of the vectors $\hat u^a$ and $\hat v^b$.  It is convenient to
expand this function in terms of Legendre polynomials:
\begin{equation} C(\gamma) = \sum_{l=2}^\infty { 2 l + 1 \over 4  \pi}
a^2_l W_l P_l(\cos \gamma).  \end{equation} The coefficients $a^2_l$
are referred to as the multipole moments of the expansion.  Note that
the monopole term ($l=0$) is absent; the dipole term ($l=1$) is
generally removed from the data because it depends mostly upon the
observer's peculiar velocity.  The quantity measured by a given
experiment is affected by the filtering properties of the optics and
receivers, which determine the angular range over which the experiment
is sensitive; the effect of this filtering is incorporated into the
``weight function" $W_l$, which differs from experiment to experiment.
These weight functions are shown in Figure 1b of reference
\cite{Critt2} for a number of different experiments; for the purposes
of this paper we will consider an ``ideal" experiment that is equally
sensitive at all angular scales and has a weight function $W_l=1$.

With cosmological models concrete enough to make definite theoretical
predictions, one may calculate the {\it expected value} of this
correlation function.  If the cosmological model is isotropic then the
correlation function  depends only on the angle $\gamma$ between the
pair of observation points even before the ``sky averaging" in
(\ref{classcorr}) is done; thus averaging is not necessary.  It is
important to note however that in the experimental case, the multipole
moments $a^2_l$ associated with the observed sky-averaged correlation
function have definite measurable values, but given a specific
theoretical model, these actual values are impossible to predict; they
depend upon our location in universe, and additonally reflect the fact
that our universe is a single realization of the statistical ensemble
whose expected values may be determined theoretically.  Hence, the
quantity determined in this paper is an expectation value; we denote
the associated expected multipole moments by $\langle a^2_l \rangle$.
This ensemble average or expectation value is equal to a the uniform
average over observers located at all spatial locations, with all
possible choices of direction on the celestial sphere.  Thus, in
principle, one could directly compare the observed  $a^2_l$ with the
expected values $\langle a^2_l \rangle$ by averaging observational data
taken from regions of the universe that are currently not in causal
contact, but this process would take many times the age of the universe
to complete.  Hence there remains a practical problem, that of
constraining the cosmological models by comparing the expected
multipole moments $\langle a^2_l \rangle$ with the observed multipoles
$a_l^2$.  This requires statistical analysis of the expected variance
in $a_l^2$; this problem of cosmic variance will not be addressed
here.  One does expect however that since the number of independent
degrees of freedom in the $l'$th multipole moment is $2l+1$, for a good
cosmological model the observed $a_l^2$ and the expected $\langle a^2_l
\rangle$ should be very closely equal for large $l$.

This paper considers the angular correlation function for models  of
the universe that pass through an early inflationary stage.  More
precisely, we consider  models where the cosmological length scale
(scale factor) undergoes a long period of exponential expansion,
characterized by a constant, positive energy density  and  a constant,
negative isotropic pressure of equal magnitude.  Such cosmological
models are attractive because they solve the horizon and flatness
problems in a ``natural'' way \cite{Starobinskyinflation,Guth}.  For
this reason an enormous variety of mechanisms for inflation have been
proposed during the past decade.

Since the proposed inflationary models differ in significant ways, they
make certain predictions that are quantitatively very different.  As an
example, the perturbations in the CBR temperature that result from
fluctuations in the matter-density are model-dependent because the
matter content of these models is limited only by the imagination of
the model-builder.  Fortunately there are certain predictions of
inflationary models that are {\it independent} of the details of the
model.  One significant example is the subject of this paper; the
perturbations of the CBR temperature that result from the gravitational
wave fluctuations.   These perturbations depend only on a single
parameter: the energy-density during the period of exponential
expansion.

In the simplest perturbed FRW models, one may classify the
perturbations which produce  fluctuations of the CBR temperature as
scalar, vector or tensor in nature.  The complete angular correlation
function $C(\gamma)$ is the sum of terms arising from each of these;
one generally assumes that these add incoherently (or in quadrature).
For a large class of ``slow rollover'' inflationary models,  the
expectation value of the angular correlation function resulting from
scalar perturbations is \begin{equation} C(\gamma) = {3 \over 2 \pi}
\langle a^2_2 \rangle \biggl( \ln {2 \over 1 - \cos \gamma} -1 - {3
\over 2} \cos \gamma \biggr).  \label{closedformscalar} \end{equation}
(Note that the dipole moment has been removed.)  This corresponds to an
expected spectrum of coefficients $\langle a^2_l \rangle$ given by
$\langle a^2_l \rangle = {6 \langle a_2^2 \rangle \over  l (l+1)}$.
Only the overall amplitude of the correlation function, here determined
by the expected value of the quadrupole moment $\langle a^2_2 \rangle
$, varies from model to model.  The correlation function due to vector
perturbations is typically very small and is neglected.  In this paper
we only consider the contribution to the angular correlation
function from the tensor (or gravitational wave)
perturbations; as we will
shortly explain, these are entirely determined by the energy-density
during the inflationary phase and are otherwise model-independent.

There is a substantial body of research on this topic.  In the
following brief review we do not include much of the important work on
the effects of scalar density fluctuations on the CBR, but principally
discuss the work on CBR fluctuations induced by gravitational wave
perturbations.  The original discovery that cosmological expansion
could create particles is due to Parker \cite{ParkerCreation} and
Zel'dovich \cite{ZeldovichCreation}.  However, they apparently assumed
that the linearized gravitational wave equation would be conformally
invariant and hence that no gravitons could be created.   This
oversight was corrected by Grishchuk \cite{GrishchukCreation} who
showed that due to the lack of conformal invariance a period of rapid
cosmological expansion could result in the non-adiabatic amplification
of weak classical gravitational waves.  The corresponding classical
process for black holes (super-radiant scattering) implies the quantum
effect (Hawking radiation).  In similar fashion, Ford and Parker
\cite{FordParker} showed how one could systematically quantize the
linearized gravitational field on an FRW background, and calculated the
spectrum of gravitons created by the cosmological expansion.

Starobinsky investigated this process in detail for inflationary
cosmologies (but before the term ``inflation" had been coined
\cite{Starobinskyinflation} and before the advantages of such a period
of expansion had been fully appreciated and explained \cite{Guth}) and
found the power spectrum of gravitational radiation that would be left
behind \cite{Starobinky1}.  The quadrupole ($l=2$) and octupole ($l=3$)
anisotropies in the CBR induced by the resulting gravitational
perturbations were later calculated by Rubakov, Sazhin and Veryaskin
\cite{Rubakov}.  While the methods used and the interpretation of the
results are entirely correct, this work suffers from technical errors.
In particular, the octupole moment is correct but the quadrupole moment
has the wrong value: the right-hand side of their equation (7) reads
``$2.4\epsilon_V /M^4_{\rm pl}$'', but the correct result is  $(\Delta
T_0/T_0)^2_{\rm quadr}=1.54\epsilon_V /M^4_{\rm pl}$.  Soon afterwards,
similar results were published by Fabbri and Pollock \cite{Fabbri}, who
gave the first general formula for the $l'$th multipole moment in
inflationary models.  This work has a minor typographical error (the
right-hand side of their equation (14) should be doubled) but otherwise
their results are correct.  About a year later, this work was repeated
and generalized for power-law inflation by Abbott and Wise
\cite{AbbotWise}, who also give a (now standard) correct formula for
the $l'$th multipole moment.  Shortly thereafter, Starobinsky
\cite{Starobinsky2} also published the results of an independent
analysis, giving the same formula for the $l'$th multipole and
correcting the errors in \cite{Rubakov} and \cite{Fabbri}.  This early
work all considered the spatially flat $k=0$ FRW models; it was
generalized to the $k= \pm 1 $ cases by Abbott and Shaefer
\cite{AbbottShaefer}, who systematically considered the CBR
fluctuations induced by all three (scalar, vector, and tensor) types of
perturbations to the $k=0,\pm 1$ FRW metrics in inflationary models.
The energy-density of the classical gravitational waves resulting from
inflation was re-examined by Abbott and Harari \cite{AbbottHarari}, who
stressed the quantum-mechanical origin of this radiation, and by Allen
\cite{Allen} who elucidated the first complete formula for the
power-spectrum in gravitational radiation, and it's connection to the
low-frequency instability (and peculiar infra-red behavior) of de
Sitter space.  As one consequence, Allen showed that the energy-density
in gravitational waves falls off more slowly with time than the
corresponding background energy density of the dust driving the FRW
expansion.  This work was subsequently extended by Ressell and Turner
\cite{RessellTurner} who examined the effect of a ``dustlike" phase
during which the scalar field oscillated and decayed on the
gravitational radiation power-spectrum.  The work was then further
generalized by Sahni \cite{Sahni}, who repeated these calculations for
power-law inflation.

Interest in this subject was re-awakened by the publication of the COBE
data \cite{SmootEtAl}.  A number of papers have examined whether the
different $l-$dependence of the scalar and tensor contributions to
$\langle a^2_l \rangle$ permit one to determine their separate
amplitudes.  Typically these compare the scalar
contributions expected from a period of
quasi-exponential (slow-roll) expansion (which
inflates any early perturbations to well beyond today's Hubble radius)
to the tensor perturbations.  These include work by Souradeep and Sahni
\cite{SouradeepSahni}, Liddle and Lyth \cite{LiddleLyth}, Davis et.
al.  \cite{DavisEtAl}, Salopek \cite{Salopek}, Lucchin, Matarrese and
Mollerach \cite{Lucchin1}, Dolgov and Silk \cite{DolgovSilk}, Turner
\cite{Turner} , and Crittenden et. al. \cite{Critt2,Crittenden}.
Another possibility is that one may distinguish the scalar and tensor
contributions to the multipole moments by examining the polarization of
the CBR.  This has been examined by Harari and Zaldarriaga
\cite{Harari}, by Crittenden, Davis and Steinhardt \cite{Crittenden},
and by Ng and Ng \cite{KaLok}.

Krauss and White \cite{KraussWhite} have used statistical methods and
the COBE data to put tighter constraints on the energy-density during
an inflationary epoch.  Further details of a Monte-Carlo simulation
were given by White \cite{White} who also presented a concise
derivation of the formula for the $\langle a^2_l \rangle$ due to tensor
perturbations, and a table of the first ten $\langle a^2_l \rangle$.
The effects of cosmic variance on the ability to distinguish the scalar
and tensor perturbations and the slope of the power-spectrum was also
considered by White, Krauss and Silk \cite{White2}.

A related analysis has  been performed by Bond et. al. \cite{Bond} and
by Crittenden et. al. \cite{Critt2} who investigate how well one can
measure a number of important cosmological parameters from the
collection of anisotropy observations.  It turns out that the first few
multipole moments are sensitive to very-long wavelength modes which
probe well outside our current Hubble radius.  Stevens, Scott and Silk
\cite{Stevens} and Starobinsky \cite{Starobinsky2} have used these
measurements to put new lower limits on the ``circumference" of the
universe, in the case where it has toroidal spatial topology.  Such
analysis may also be possible in the spatially open case, where the
Sachs-Wolfe effect has been studied by Ratra and Peebles \cite{Ratra}.

Grishchuk has also examined the multipole moments arising from
gravitational wave perturbations \cite{Grishchuk1,Grishchuk2}, adapting
the terminology and techniques of quantum optics to carry out the
analysis.  Grischchuk stresses the importance of the phase correlations
between the modes of the metric perturbations; we agree with this
conclusion but do not use Grischchuk's ``squeezed state" representation
of the field operator.  It is possible to obtain identical results
using only the standard formalism of curved-space quantum field theory
developed in \cite{FordParker}.  Our conclusion is that the standard
ansatz for the so-called ``Gaussian spectrum of initial perturbations''
only gives reliable results for small values of $l$; for the higher
moments the phase relationship between the positive- and
negative-frequency components of the wave-functions do affect the
multipole moments.

Deviations from Gaussian behavior may in principle be observed through
the three-point angular correlation function.  This was first
calculated by Falk, Rangarajan and Srednicki \cite{Falk}; the
implications of these results and further analysis have been carried
out by Luo and Schramm \cite{Schramm} and Srednicki \cite{Srednicki}.

The physical processes giving rise to the CBR temperature fluctuations
may be understood (and explained) in several ways.  We repeat the
interpretation given by Allen \cite{Allen}, which also sheds light on
our technical methods.  The period of exponential expansion is an
unstable one, from the global point of view.  During this expansion,
perturbations of the spatial geometry tend to freeze in dimensionless
amplitude, so that when viewed globally the spatial sections become
more and more distorted.  However, another consequence of the rapid
expansion is that locally, any observer can only see (within her Hubble
radius) a smaller and smaller region of
this spatial section.  Hence from the observer's
local point of view, the spacetime is getting
closer and closer to a perturbation-free de Sitter spacetime.  One
consequence of this global instability/local stability is that
gravitational perturbations which are of local origin (for example, due
to thermal fluctuations) are very rapidly redshifted in wavelength and
amplitude.  At late times, after sufficient inflation, these
perturbations are no longer visible to an observer; the only
perturbations which remain visible are those of quantum origin (the
zero-point fluctuations associated with the uncertainty principle)
because these fluctuations extend up to arbitrarily high frequency and
can not be redshifted away.  (In similar fashion, the quanta radiated
by an evaporating black hole at late times are due to quantum
zero-point fluctuations at very high frequency close to the horizon.)
Hence, to determine the gravitational perturbations present at late
times, we assume that the initial state of the universe was the vacuum
state appropriate to de Sitter space, containing {\it only} the quantum
fluctuations and {\it no} additional excitations.  For this reason, one
can do a calculation based entirely on ``first principles"; the
amplitude of the primordial fluctuations follows directly from the
canonical commutation relations obeyed by the linearized gravitational
field, or in physical terms, directly from the uncertainty principle.
The ``particle production" in this case is the production of pairs of
gravitons, whose collective effects (since the occupation numbers are
large, and they are bosons) appear as classical gravitational
radiation.  Thus, we determine the expected value of the correlation
function (\ref{classcorr}) by finding the expectation value of ${\delta
T \over T}(\hat u^a) {\delta T \over T}(\hat v^b)$ in the de Sitter
vacuum state.

The published calculations have two shortcomings which are addressed in
the present work.  The first is pedagogic.   The calculations which
have been published are all rather sketchy; to reproduce the results
requires many pages of calculation which are not given in full but are
left as an exercise to the reader.  We believe that our method of
performing this calculation is new; it is short enough and elegant
enough so that {\it all} of the details can be shown explicitly.  The
second advantage is quantitative.  The previously published
calculations are made using a ``long wavelength" approximation to the
mode functions, which is accurate for determining the effects of the
lowest multipoles, but inaccurate for the higher ones.  It is not
obvious from the published work how to remove this approximation to
obtain more accurate results; in the present work we give exact
expressions for the correlation function.  The shortcomings of the
standard ``long wavelength" approximations have been pointed out in the
recent work of Turner, White and Lidsey \cite{Turner2}, who use
numerical methods to integrate the wave equation for the mode functions
and who obtain results that appear identical to our exact formula.

This paper is organized as follows.   After a few notes on notation,
section \ref{section2}  begins with the classic formula for the
Sachs-Wolfe effect in spatially flat FRW cosmological models.  This is
used to derive an expression for the correlation function $C(\gamma)$
due to the gravitational radiation, under the assumption that the
initial state of the universe is a vacuum state with only zero-point
perturbations.  In section \ref{section3} we derive from first
principles the normalization condition on the graviton wave functions.
Section \ref{section4} describes a simple inflationary cosmological
model, also used in \cite{Allen}.  In this model, the universe
``begins" with an infinite period of inflation, then makes an
instantaneous transition to a radiation-dominated stage, and then later
makes another instantaneous transition to a matter-dominated stage.  We
then find the normalized graviton wave functions appropriate to that
model (and the corresponding Bogolubov coefficients). We also show how
the standard results appear as a low-frequency approximation to the
exact expressions.  Section \ref{section5} is an attempt to obtain a
closed form for $C(\gamma)$; this attempt does not succeed  but some
progress is made.  Section \ref{section6} compares the results of our
exact expression for the expansion coefficients $\langle a^2_l \rangle$
with the more standard results, and includes a discussion of some
recent literature on the subject.  Because the high-frequency modes
affect the temperature perturbations on small angular scales, the exact
$\langle a^2_l \rangle$ agree with those given by the standard
approximations for small $l$, and are different for large $l$.  Finally
a pair of appendices show an alternative derivation of the formulae
contained in section \ref{section2}, and contain a brief description of
the numerical techniques used in section \ref{section6}.

Throughout this paper, we use units where the speed of light $c=1$.
However for clarity we have retained Newton's gravitational constant
$G$ and Planck's constant $\hbar$ explicitly.


\section{THE SACHS-WOLFE EFFECT \protect\\
AND THE ANGULAR
CORRELATION FUNCTION}
\label{section2}

\subsection{Notes on Notation}
\label{subsec:notation}

We begin with a few notes on notation. The vectors and tensors in this
section are purely spatial; they have no time components, although they
may be time-dependent functions.  In a spatially-flat FRW model, the
spatial geometry is flat Euclidean space.  Since the tensors and
vectors are spatial we raise and lower tangent space indices with the
spatial part of the conformal metric, which is just the Euclidean
metric of $\Re^3$.  In Cartesian coordinates, this is \begin{equation}
\delta_{ab}={\rm diag}(1,1,1).  \end{equation} We denote  spatial
vectors by $k^a, v^a,$ or $u^a$, and spatial tensors by  $h_{ab},
e_{ab}$, or $P_{ab}$.  The Latin indices $a,b,\ldots ,f$ run from 1 to
3.  Associated with any spatial vector is its magnitude, denoted by the
vector symbol without a tangent space index. For example, the magnitude
of the vector $k^c$ is denoted $k$, where \begin{equation}
k\equiv\sqrt{k^a k_a}=\sqrt{\delta_{ab}k^a k^b}.  \end{equation} A
special notation is used for spatial vectors with unit magnitude.  The
unit spatial vector $\hat k^a$ is defined by \begin{equation} \hat
k^a\equiv {k^a\over k}, \end{equation} so that $\hat k^a\hat k_a=1$.
Thus one may decompose any spatial vector $u^a $ into a magnitude and a
unit vector, and express it  as \begin{equation} u^a=u\hat u^a.
\end{equation} We use this notation throughout this section.

Often we  need to integrate over all possible magnitudes and
orientations of a spatial vector.  To integrate over $k^a$, we write
\begin{equation} \int d^3k=\int_0^\infty dk\>k^2\int_0^\pi d\theta_k
\sin\theta_k\int_0^{2\pi}d\phi_k\equiv \int_0^\infty dk\>k^2\int
d\Omega_{\hat k}, \end{equation} denoting the polar angle associated
with $k^c$ by $\theta_k$, and the azimuthal angle by $\phi_k$.  In a
similar way we will denote a function of the polar and azimuthal angles
(such as a spherical harmomic function) as \begin{equation}
Y_{lm}(\theta_k,\phi_k)\equiv Y_{lm}(\hat k^c).  \end{equation} For
example,  the orthonormality condition for the spherical harmonics is
\begin{equation} \int d\Omega_{\hat k} Y_{lm}(\hat k^c) Y^*_{pq}(\hat
k^c)=\delta_{lp}\delta_{mq}.  \end{equation} Note that we never
integrate over the polar and azimuthal angles separately.

Hilbert space operators are denoted by an overbar, for example
\begin{equation} \bar a|\psi\rangle=|\psi'\rangle, \end{equation} and a
$\dagger$ denotes the adjoint operator.  A $*$ denotes complex
conjugation.

\subsection{The Sachs-Wolfe Effect}
\label{classSW}

In a perfectly isotropic universe the CBR would  have the same
temperature in all directions on the celestial sphere.  If, however,
the cosmological metric is perturbed away from isotropy, the
temperature observed today fluctuates over the celestial sphere, even
if the last-scattering surface had uniform  temperature.  The
Sachs-Wolfe formula \cite{SachsWolfe} expresses the  temperature
fluctuations in terms of the metric perturbations, to lowest order in
the perturbation.

Consider a spatially-flat FRW universe perturbed away from isotropy.
The perturbed metric in comoving gauge is \begin{equation}
ds^2=a^2(\eta)[-d\eta^2+\big(\delta_{ab}+h_{ab}(\eta,x^c )\big) dx^a
dx^b], \label{ourmetric} \end{equation} where $\eta$ is the conformal
time, and $a(\eta)$ is the  scale factor. Imagine a single photon {\it
emitted} at conformal time $\eta={\eta_{\rm e}}$ and {\it observed} at
conformal time $\eta={\eta_{\rm obs}}$.  As the photon propagates
through the spacetime the metric perturbations $a^2(\eta)h_{ab}$
perturb the null path of the photon.  One may choose spatial
coordinates so that (to zero order in $h_{ab}$) the photon follows a
radial path in space on its way to the observer, who is located at the
origin $x^a=0$. Furthermore, one may parametrize the path of the photon
by $\lambda$, so that the spatial path of the photon is
\begin{equation} x^a(\lambda)=D(\lambda)\hat u^a, \end{equation} where
\begin{equation} D(\lambda)=({\eta_{\rm obs}}-{\eta_{\rm e}}-\lambda),
\label{Ddefinition} \end{equation} $\hat u^a$ is a unit vector pointing
radially out from the origin, and $\lambda$ varies from ${\lambda_{\rm
e}} $ to ${\lambda_{\rm obs}} $ with \begin{eqnarray} {\lambda_{\rm e}}
&=&0,\\ {\lambda_{\rm obs}} &=&{\eta_{\rm obs}}-{\eta_{\rm
e}}.\\ \nonumber \end{eqnarray} Sachs and Wolfe have shown that to
first order in $h_{ab}$ the observed redshift of the photon is given by
\begin{equation} 1+Z={a({\eta_{\rm obs}})\over a({\eta_{\rm
e}})}\bigg(1+{1\over 2}\int^{{\lambda_{\rm obs}} }_{{\lambda_{\rm e}} }
\hat u^a\hat u^b\bigg[{\partial \over \partial\eta}h_{ab} (\eta,
D(\lambda)\hat u^c)\bigg]_{\eta ={\eta_{\rm e}}+\lambda}
d\lambda\bigg).  \label{redshift} \end{equation} This equation is
equivalent to (39) in reference \cite{SachsWolfe} for the specialized
case of gravitational-wave perturbations.

The CBR is an ensemble of many  photons which were last scattered at
conformal time $\eta={\eta_{\rm e}}$ by the primordial plasma of ionized
hydrogen
and electrons. Using (\ref{redshift}) one obtains the temperature
fluctuation $\delta T$ of the CBR measured at the point on the
celestial sphere pointed to by the unit vector $\hat u^c$:
\begin{equation} {\delta T\over T}(\hat u^c) ={1\over
2}\int^{{\lambda_{\rm obs}} }_{{\lambda_{\rm e}} }\hat u^a\hat
u^b\bigg[{\partial
\over\partial\eta}h_{ab}(\eta, D(\lambda)\hat u^c)
\bigg]_{\eta={\eta_{\rm e}}+\lambda}d\lambda.  \label{swformula} \end{equation}
This formula embodies the  Sachs-Wolfe effect, and is equivalent to
(42) in reference \cite{SachsWolfe} for the special case of
gravitational-wave perturbations.

\subsection{The Metric Perturbation $h_{ab}$}
\label{sec: hab}

As noted in the introduction, we examine the transverse, traceless,
tensor part of the metric perturbation in models of the universe that
pass through an early inflationary stage.  The period of exponential
inflation is unstable, and as a result of the rapid expansion,
perturbations of the spatial geometry freeze in dimensionless
amplitude. From any observer's local point of view the spacetime
quickly approaches a perturbation-free de Sitter spacetime. At late
times perturbations of local origin are extremely redshifted in both
wavelength and amplitude, leaving only perturbations of quantum origin
(zero-point fluctuations) as the significant contribution to the tensor
part of the metric perturbations. For this reason we assume that the
initial state of the universe is the de Sitter space vacuum state
containing only quantum fluctuations.

Since the significant tensor perturbations are quantum in origin, we
replace the classical metric perturbation $h_{ab}$ in (\ref{swformula})
by the  Hilbert space operator $\bar h_{ab}$  appropriate for the
linearized theory of gravity. The plane wave expansion of $\bar h_{ab}$
is \begin{eqnarray} \bar h_{ab}(\eta,  x^c)=\int d^3k\> \bigg(& &e^{i
k^d x_d}\big[e_{ab}( k^c)\phi_R(\eta,k^c) \bar a_R( k^c)
+e^*_{ab}(k^c)\phi_L(\eta,k^c)\bar a_L( k^c) \big]\nonumber\\
 &&+ e^{-i  k^d x_d}\big[e^*_{ab}(k^c) \phi_R^*(\eta,k^c) \bar
a^\dagger_R( k^c) +e_{ab}(k^c) \phi_L^*(\eta,k^c)\bar a^\dagger_L( k^c)
\big]\bigg).  \label{barhab} \end{eqnarray} Here $\bar a_R(k^c)$ and
$\bar a_L(k^c)$ (their Hermitian conjugates) are annihilation
(creation) operators that destroy (create) a right  or left circularly
polarized graviton. These operators obey the commutation relations
\begin{equation} [\bar a_L(  k^a),\bar a_L^\dagger( k'{}^a)]=[\bar
a_R(  k^a), \bar a_R^\dagger(k'{}^a)]=\delta^3(k^a-k'{}^a),
\label{acommutation} \end{equation} with all other commutators
vanishing.  The graviton mode functions for the left and right
polarizations are $\phi_L(\eta,k^c)$ and $\phi_R(\eta,k^c)$
respectively. If the spacetime is isotropic and homogeneous, and
therefore does not single out any preferred directions, one may choose
a particle basis so that the mode functions depend on the magnitude $k$
only.  One may also choose a particle basis that does not distinguish
between the two possible spatial orientations, so that the left- and
right-handed gravitons have the same mode functions. One then has
\begin{equation} \phi_L(\eta,k^c)=\phi_R(\eta,k^c)\equiv\phi(\eta,k).
\end{equation} This mode function $\phi(\eta,k)$ obeys the massless
Klein-Gordon equation \cite{FordParker} \begin{equation}
\ddot\phi+2{\dot a(\eta)\over a(\eta)}\dot\phi+k^2\phi=0,
\label{KleinGordon} \end{equation} where $a(\eta)$ is the cosmic scale
factor, and \begin{equation} \cdot\equiv{\partial\over\partial\eta} .
\end{equation} If one demands that  $\bar h_{ab}$ obey  canonical
commutation relations, the  commutation relations (\ref{acommutation})
imply that the mode function satisfy normalization conditions. The
normalization condition is defined in (\ref{phinormalization}).

The  tensors $e_{ab}( k^c)$ and $e^*_{ab}(k^c)$ in the expansion
(\ref{barhab}) are the polarization tensors for a circularly polarized
basis.  We first define the so called plus $(+)$ and cross $(\times)$
polarizations.  Consider a mode or wave propagating in the  $\hat k^c$
direction.  One may define two unit length vectors $\hat m^c$ and $\hat
n^c$  orthoganol to $\hat k^c$, and orthoganol to each other, so that
the set $(\hat k^c,\hat m^c,\hat n^c)$ is a right-handed triad with
\begin{equation} \hat k^a\hat m_a=\hat k^a\hat n_a=\hat m^a\hat n_a=0.
\label{orthog} \end{equation} In terms of these unit vectors the plus
and cross polarizations are defined as \begin{eqnarray}
e^{(+)}_{ab}(k^c)&=&\hat m_a(k^c)\hat m_b(k^c)- \hat n_a(k^c)\hat
n_b(k^c)\\ e^{(\times)}_{ab}(k^c)&=&\hat m_a(k^c)\hat n_b(k^c)+ \hat
n_a(k^c)\hat m_b(k^c).  \label{pluscross} \end{eqnarray} The plus and
cross polarization tensors together form a complete basis for the
tensor (spin 2) perturbations \cite{SachsWolfe}.  Note that both the
plus and cross polarizations are transverse, traceless, and symmetric:
\begin{equation} e^{({\buildrel +\over\times} )}_{ab}(k^c)
k^a=e^{({\buildrel +\over\times})}{}^a{}_a(k^c) =e^{({\buildrel
+\over\times})}_{[ab]}(k^c)=0.  \end{equation} One may  define the
circular polarization tensor $e_{ab}(k^c)$ in terms of the plus and
cross polarizations as \begin{eqnarray} e_{ab}(k^c)&=&{1\over\sqrt
2}\big[e^{(+)}_{ab}(k^c)
+ie^{(\times)}_{ab}(k^c)\big],\label{circpol}\\ &=&{1\over\sqrt 2}
\big[\hat m_a(k^c)+i \hat n_a(k^c)\big]\big[\hat m_b(k^c) +i \hat
n_b(k^c)\big].  \label{circpolar} \end{eqnarray} The polarization
tensor $e^*_{ab}(k^c)$ is just the complex conjugate of the
polarization tensor (\ref{circpol}).  The tensors $e_{ab}(k^c)$ and
$e^*_{ab}(k^c)$ also form a complete basis for the tensor (spin 2)
perturbation.

The vectors $\hat m_c(k^c)$ and $\hat n_c(k^c)$ are not unique. Any two
unit vectors that satisfy (\ref{orthog}) may be used to define the
polarization tensors. Any other right-handed triad of vectors such as
$(\hat k^c,\hat m'^c,\hat n'^c)$, however, can be obtained by rotating
the triad $(\hat k^c,\hat m^c,\hat n^c)$ through an angle $\varphi$
about $\hat k^c$. Under this rotation, \begin{equation}
e'_{ab}(k^c)=e^{-2i\varphi}e_{ab}(k^c).  \end{equation} This shows that
gravitons are a spin 2 field, since the spin (or more precisely, the
helicity) of a field is defined as the number of times the phase of the
field changes by $2\pi$, when the coordinate system is rotated once
around the momentum vector of the field.

The polarization tensors are closely related to the  tensor that
projects onto a sphere of radius $k$, at a point $k^c$ . We define the
projection tensor $P_{ab}( k^c)$ by \begin{equation} P_{ab}(
k^c)\equiv\delta_{ab}-\hat k_a\hat k_b , \label{pab} \end{equation} so
that \begin{equation} P_{ab}\hat k^a=P_{ab}\hat k^b=0,\>\> {\rm
and}\>\> P_{ab}P^b{}_c=P_{ac}.  \end{equation} This tensor projects
onto the two-surface orthoganol to $k^c$, which is just the two-sphere
of radius $k$.  To relate the projection tensor $P_{ab}$ to the
polarization tensors, consider the Euclidean metric $\delta_{ab}$ on
$\Re^3$.  One may express $\delta_{ab}$ using  the three unit vectors
$\hat k^c,\hat m^c$, and $\hat n^c$:  \begin{equation} \delta_{ab}=\hat
k_a\hat k_b+\hat m_a\hat m_b +\hat n_a\hat n_b.  \label{pabandmn}
\end{equation} Using (\ref{pab}) and (\ref{pabandmn}) one may write the
projection tensor as \begin{equation} P_{ab}(k^c)=\hat m_a(k^c)\hat
m_b(k^c)+ \hat n_a(k^c)\hat n_b(k^c).
\label{pinmn} \end{equation} From (\ref{circpolar}) and (\ref{pinmn})
one may quickly verify the identity \begin{equation}
e_{ab}(k^e)e^*_{cd}(k^e)+
e^*_{ab}(k^e)e_{cd}(k^e)=P_{ac}(k^e)P_{bd}(k^e)+
P_{ad}(k^e)P_{bc}(k^e)-P_{ab}(k^e)P_{cd}(k^e) .  \label{identity}
\end{equation} Later  we  use this identity to find an elegant
expression for the angular correlation function.

\subsection{The Two-Sphere of Radius $k$}
\label{sec:sphere}

Besides being the projection tensor onto the two-sphere of radius $k$,
$P_{ab}$ is the natural metric induced on this two-surface by the flat
metric on $\Re^3$. Since the two-sphere is a maximally symmetric
two-manifold, one may immediately write the Riemann tensor on this
two-surface as \begin{equation} \tilde R_{abcd} ={2\over k^2}
 P_{a[c}P_{d]b}.  \label{riemann} \end{equation} The factor of $k^{-2}$
appears because the two-sphere has radius $k$. We denote the covariant
derivative on  this surface by $\tilde\nabla_a$, and define the
Laplacian ${\tilde{\smash{\Box}\vphantom{\nabla}}}$ on this surface by
\begin{equation}
P^{ab}\tilde\nabla_a\tilde\nabla_b=
\tilde\nabla^b\tilde\nabla_b\equiv{\tilde{\smash{\Box}\vphantom{\nabla}}}.
\label{laplacian}
\end{equation} The spherical harmonics are eigenfunctions of this
Laplacian, and obey the eigenfunction equation \begin{equation}
 {\tilde{\smash{\Box}\vphantom{\nabla}}} Y_{lm}(\hat k^c)=-{l(l+1)\over k^2}
Y_{lm}(\hat k^c) .
 \label{eigen} \end{equation} Again the factor of $k^{-2}$ appears
because the  the two-sphere has radius $k$.

Using the definition of the Riemann tensor, the identity
(\ref{riemann}), and the eigenfunction equation (\ref{eigen}), we can
derive a useful identity:  \begin{eqnarray}
 {\tilde{\smash{\Box}\vphantom{\nabla}}}{\tilde\nabla}^a
 Y^*_{lm}&=&{\tilde\nabla}^b{\tilde\nabla}^a{\tilde\nabla}_b
 Y^*_{lm}\nonumber\\
&=&\big({\tilde\nabla}^b{\tilde\nabla}^a{\tilde\nabla}_b
-{\tilde\nabla}^a{\tilde\nabla}^b{\tilde\nabla}_b
+{\tilde\nabla}^a{\tilde\nabla}^b{\tilde\nabla}_b\big)
Y^*_{lm}\nonumber\\ &=&\tilde R^{ba}{}_{bc}{\tilde\nabla}^c Y^*_{lm}
+{\tilde\nabla}^a {\tilde{\smash{\Box}\vphantom{\nabla}}}
Y^*_{lm}\nonumber\\ &=&{1\over k^2}{\tilde\nabla}^a
Y^*_{lm}+{\tilde\nabla}^a {\tilde{\smash{\Box}\vphantom{\nabla}}}
Y^*_{lm}\nonumber\\ &=&\bigg[{-l(l+1)+1\over k^2}\bigg]
{\tilde\nabla}^a Y^*_{lm}
.  \label{yident} \end{eqnarray} This formula will prove useful in our
derivation of the angular correlation function.

\subsection{The Angular Correlation Function}
\label{sec:acf}

\subsubsection{The Sachs-Wolfe Operator}

The Sachs-Wolfe formula (\ref{swformula}) is a result from classical
general relativity, giving the temperature fluctuations of the CBR over
the celestial sphere as a function of metric perturbations.  As noted
above, however, in inflationary models the surviving metric
perturbations are quantum in origin; without further justification we
replace the classical metric perturbation $h_{ab}$ in the standard
Sachs-Wolfe formula (\ref{swformula}) by the quantum field operator
$\bar h_{ab}$.  The temperature fluctuation at a point on the celestial
sphere is now a Hilbert space operator, given by \begin{equation}
{\overline{\delta T}\over T}(\hat u^c) ={1\over 2}\int^{{\lambda_{\rm
obs}} }_{{\lambda_{\rm e}} }\bigg[\hat u^a\hat u^b{\partial
\over\partial\eta}\bar h_{ab}(\eta,D(\lambda)\hat u^c)
\bigg]_{\eta={\eta_{\rm e}}+\lambda}d\lambda .  \label{qsw}
\end{equation} We will refer to (\ref{qsw}) as the Sachs-Wolfe
operator.

Since the  Sachs-Wolfe operator is parametrized by coordinates on the
celestial two-sphere, it is natural to decompose it into an expansion
of (normalized) spherical harmonics on the two-sphere. Using the
orthogonality of the spherical harmonics, one can write the Sachs-Wolfe
operator as \begin{equation} {\overline{\delta T}\over T}(\hat
u^a)=\sum_{lm} \bar C_{lm} Y_{lm}(\hat u^a) , \label{expan}
\end{equation} where the expansion coefficient operator $\bar C_{lm}$
is \begin{equation} {\bar C}_{lm}={1\over 2}\int^ {{\lambda_{\rm obs}}
}_{{\lambda_{\rm e}} }d\lambda\>\int d\Omega_{\hat u} \> \hat u^a\hat
u^b Y^*_{lm}(\hat u^c)\bigg[ {\partial
\over\partial\eta}\bar{h}_{ab}(\eta, D(\lambda)\hat u^c)
\bigg]_{\eta=\eta_e+\lambda} .  \label{clm}
 \end{equation} For the metric perturbation operator $\bar h_{ab}$ we
use (\ref{barhab}). Since the {\it first derivative} of the metric
perturbation, not the perturbation itself, appears in the expression
for the expansion coefficient $\bar C_{lm}$, it is useful to define the
dimensionless function \begin{equation} F(\lambda,k)\equiv
k^{\scriptstyle 1/2}\bigg[
{\partial\over\partial\eta}\phi(\eta,k)\bigg]_{\eta= {\eta_{\rm
e}}+\lambda} .  \label{Fdefinition} \end{equation} Then from
(\ref{barhab}) and  (\ref{clm}) we obtain for the expansion coefficient
operator \begin{eqnarray} \bar C_{lm}&=&{1\over 2}\int\limits^
{{\lambda_{\rm obs}} }_{{\lambda_{\rm e}} }d\lambda\int  d\Omega_{\hat
u}\int {d^3 k\over k^{1/2}} \> \hat u^a\hat u^b\>Y^*_{lm}(\hat
u^c)\nonumber\\ &&\times\bigg\{ e^{iD(\lambda) k^d\hat u_d}
F(\lambda,k)\big[  e_{ab}(k^c)\bar a_R(  k^c) + e^*_{ab}(k^c)\bar
a_L(k^c)\big]\nonumber\\ && +e^{-iD(\lambda) k^d\hat u_d}
F^*(\lambda,k)\big[  e^*_{ab}(k^c)\bar a^\dagger_R(  k^c) +
e_{ab}(k^c)\bar a^\dagger_L(k^c)\big] \bigg\} .  \label{bigclm}
\end{eqnarray} We use this expansion of the Sachs-Wolfe operator to
examine the angular correlation function.

\subsubsection{Angular Correlation Function $ C(\hat v^a,\hat u^a)$}
\label{section2subsection2}

The quantity of interest is the angular correlation  function
(\ref{classcorr}). Since the temperature fluctuations are now
represented by a Hilbert space operator, the angular correlation
function is a matrix element:  \begin{equation} C(\hat v^a,\hat
u^a)\equiv\bigg\langle 0\bigg| \bigg({\overline{\delta T} \over
T}\bigg)^\dagger(\hat v^a){\overline{\delta T}\over T}(\hat
u^a)\bigg|0\bigg\rangle .  \label{qcorr} \end{equation} Here the
quantum state $|0\rangle$ is the initial quantum state of the universe,
which we have taken to be the de Sitter space vacuum state  for reasons
discussed both in the introduction and in section \ref{sec: hab}.

Based on the isotropy of the FRW model and of the state $|0\rangle$,
one expects the correlation function to be rotationally invariant; i.e.
to depend only on the angle $\gamma$ where $\cos\gamma\equiv\hat
v^c\hat u_c$.  Using the expansion (\ref{expan}), one may express the
angular correlation function in the form \begin{equation} C(\hat
v^a,\hat u^a)=\sum_{lm}\sum_{pq} \langle 0|{\bar C}_{pq}^\dagger \bar
C_{lm}|0\rangle Y_{lm}(\hat u^a)Y_{pq}^*(\hat v^a).  \label{cexpan}
\end{equation} Since one expects the correlation function to be
rotationally invariant, one ought to be able to write the matrix
element ${\langle 0|{\bar C}_{pq}^\dagger \bar C_{lm}|0\rangle}$ as
\begin{equation} {\langle 0|{\bar C}_{pq}^\dagger \bar
C_{lm}|0\rangle}=\langle a_l^2\rangle\delta_{lp}\delta_{mq},
\label{shortmatrixelement} \end{equation} and then use (\ref{bigclm})
for $\bar C_{lm}$ and solve for $\langle a^2_l\rangle$.  In appendix
\ref{appendixA} we make this assumption, and obtain $\langle
a_l^2\rangle$ somewhat more directly.

For now, however, we show by direct calculation that the correlation
function is rotationally invariant.  Using  (\ref{bigclm}) the matrix
element ${\langle 0|{\bar C}_{pq}^\dagger \bar C_{lm}|0\rangle}$ is
\begin{eqnarray} &&{\langle 0|{\bar C}_{pq}^\dagger \bar
C_{lm}|0\rangle}={1\over 4}\int\limits^ {{\lambda_{\rm obs}}
}_{{\lambda_{\rm e}}  }d\lambda '\int\limits^ {{\lambda_{\rm obs}}
}_{{\lambda_{\rm e}} }d\lambda \int {d^3k'\over k'^{1/2}} \int
{d^3k\over k^{1/2}}F(\lambda ',k')F^{*}(\lambda,k)\nonumber\\ &&\times
\bigg[e_{ab}(\hat k'{}^e) e^{*}_{cd}(\hat k^e)\langle 0|\bar
a_R(k'{}^e) \bar a_R^\dagger(k^e)|0\rangle +e^{*}_{ab}(\hat k'{}^e)
e_{cd}(\hat k^e)\langle 0|\bar a_L(k'{}^e)\bar
a_L^\dagger(k^e)|0\rangle \bigg]\nonumber\\ & &\times \int
d\Omega_{\hat v}\int d\Omega_{ \hat u} Y_{pq} (\hat v^e)Y^*_{lm} (\hat
u^e)\hat v ^a\hat v^b\hat u^c\hat u^d
 e^{-i k^f(D(\lambda)\hat u_f-D(\lambda')\hat v_f)} .  \label{mel}
\end{eqnarray} One may immediately evaluate the two matrix elements on
the right-hand side  using the commutation relations
(\ref{acommutation}) for the creation and annihilation operators. Both
matrix elements yield the Dirac delta function $\delta^3 (k^c-k'^c)$.
Using the identity (\ref{identity}) for the polarization tensors, one
finds \begin{equation} {\langle 0|{\bar C}_{pq}^\dagger \bar
C_{lm}|0\rangle}={1\over 4} \int^ {{\lambda_{\rm obs}} }_{{\lambda_{\rm
e}}  }d\lambda '\int^ {{\lambda_{\rm obs}} }_{{\lambda_{\rm e}}
}d\lambda \int_0^\infty dk\> k\> F(\lambda ',k) F^{*}(\lambda,k)
A_{lmpq}(k,D(\lambda),D(\lambda')) , \label{cwitha} \end{equation}
where \begin{eqnarray} A_{lmpq}(k,r,r')\equiv\int d\Omega_{\hat
k}\bigg\{&&\big[ P_{ac}(k\hat k^e)P_{bd}(k\hat k^e)+ P_{ad}(k\hat
k^e)P_{bc}(k\hat k^e)-P_{ab}(k\hat k^e)P_{cd}(k\hat k^e)
\big]\nonumber\\ \times&& \psi^{cd}_{\{pq\}}(r'k\hat k^e)
\psi^*{}^{ab}_{\{lm\}}(r k\hat k^e)\bigg\} , \label{adefinition}
\end{eqnarray} and \begin{equation} \psi^{ab}_{\{lm\}}(k^c)\equiv\int
 d\Omega_{\hat u}\>\>Y_{lm}(\hat u^c)\hat u^a \hat u^b e^{i k^d \hat
u_d} .  \label{psidefinition} \end{equation} The braces in the equation
above are  to remind the reader that   $l$ and $m$ are not tangent
space indices.  We show in the next section that $A_{lmpq}(k,r,r')$ is
proportional to the Kronecker deltas $\delta_{lp} \delta_{mq}$ and is
independent of $m$, so that the correlation function is indeed
rotationally invariant.

\subsubsection{A Closed Form Expression for $A_{lmpq}(k,r,r')$}

The function $\psi^{ab}_{\{lm\}}(k^c)$ can be expressed in a way which
allows one to exploit the projection tensors  in (\ref{adefinition}).
Note from  (\ref{psidefinition}) that \begin{equation}
\psi^{ab}_{\{lm\}}(k^c)= -\nabla^a\nabla^b \int d\Omega_{\hat
u}\>\>Y_{lm}(\hat u^c)
 e^{i k^d \hat u_d} , \end{equation} where the derivative  $\nabla_a$
in Cartesian coordinates is \begin{equation}
\nabla_a\equiv{\partial\over\partial k^a}.  \end{equation} The plane
wave $e^{i k^d \hat u_d}$ can be expanded as an infinite sum of
spherical Bessel functions  $j_l(k)$ and spherical harmonics
\cite{Jackson} so that \begin{equation} \psi^{ab}_{\{lm\}}(k^c)=
-\nabla^a\nabla^b \int d\Omega_{\hat u}\>\>Y_{lm}(\hat u^c)
\bigg[4\pi\sum_{p=0}^{\infty} i^p j_p(k)\sum_{q=-p}^{p} Y^{*}_{pq}(\hat
u^c)
 Y_{pq}(\hat k^c)\bigg].  \end{equation} Using the orthonormality of
the spherical harmonics one obtains \begin{equation}
\psi^{ab}_{\{lm\}}(k^c)=-4\pi i^l\nabla^a\nabla^b j_l(k) Y_{lm}(\hat
k^c).  \end{equation} Note the dependence of the right-hand side on the
vector $k^c$; the spherical Bessel function depends only on the
magnitude $k$, and the spherical harmonic depends only on the polar and
azimuthal angles.  With this form for $\psi^{ab}_{\{lm\}}(k^c)$
(\ref{adefinition}) becomes \begin{eqnarray} A_{lmpq}(k,r,r')&&=
{16\pi^2 i^p (-i)^l\over r^2 r'^2}\nonumber\\ &&\times\int
d\Omega_{\hat k}\bigg\{ \big[ P^{ac}(k\hat k^e)P^{bd}(k\hat k^e)+
P^{ad}(k\hat k^e)P^{bc}(k\hat k^e) -P^{ab}(k\hat k^e)P^{cd}(k\hat
k^e)\big]\nonumber\\ &&\times\big[\nabla_c\nabla_d j_p(kr')Y_{pq} (\hat
k^e)\big]\big[\nabla_a\nabla_b j_l(kr)Y^*_{lm} (\hat k^e)\big]\bigg\}
.  \label{almpq} \end{eqnarray} We can now use the projection operators
to make the final integration almost trivial.

Consider how the projection tensor $P_{ab}$ acts on the gradient
$\nabla_a f(k^c)$.  The gradient in general has components both
parallel and orthoganol to $k^c$. When contracted with the gradient the
projection tensor annihilates the components parallel to $k^c$ . The
remaining components of the gradient lie entirely on the two-sphere of
radius $k$, so \begin{equation} P^a{}_b\nabla_a f(k^c)=\tilde\nabla_b
f(k^c), \label{tdel} \end{equation} where  $\tilde\nabla_a$ is the same
derivative on the two-sphere defined in section \ref{sec:sphere}.
Using  (\ref{tdel}), and noting that the spherical Bessel functions
depend only on the magnitude $k$, and are constant on the two-sphere of
radius $k$, one obtains \begin{eqnarray} A_{lmpq}(k,r,r')=&& {16\pi^2
i^p (-i)^l\over r^2 r'^2}j_p(kr')j_l(kr)\nonumber\\ \times&&\int
d\Omega_{\hat k}  \big[ 2\big( {\tilde\nabla}_a{\tilde\nabla}_b
Y_{pq}\big)\big({\tilde\nabla}^a{\tilde\nabla}^b Y^*_{lm}\big) -\big(
{\tilde{\smash{\Box}\vphantom{\nabla}}} Y_{pq}\big)\big(
{\tilde{\smash{\Box}\vphantom{\nabla}}} Y^*_{lm}\big) \big],
\end{eqnarray} where ${\tilde{\smash{\Box}\vphantom{\nabla}}}$ is the
same Laplacian on the two-sphere of radius $k$ defined in
(\ref{laplacian}).  The integrand is just derivatives on the two-sphere
of spherical harmonics, which, as discussed in section
\ref{sec:sphere}, are eigenfunctions of the Laplacian
${\tilde{\smash{\Box}\vphantom{\nabla}}}$.

The first integral on the right-hand side above can be integrated by
parts, and the second by inspection.  To help us evaluate the
integrals, we write \begin{equation} A_{lmpq}(k,r,r')= {16\pi^2 i^p
(-i)^l\over r^2 r'^2}j_p(kr')j_l(kr) \big[2 Q^{(2)}_{lmpq}(k)-
Q^{(1)}_{lmpq}(k)\big], \label{awithq} \end{equation} where
\begin{equation} Q^{(1)}_{lmpq}(k)\equiv\int d\Omega_{\hat k} \big(
{\tilde{\smash{\Box}\vphantom{\nabla}}} Y_{pq}\big)\big(
{\tilde{\smash{\Box}\vphantom{\nabla}}} Y^*_{lm}\big)= {l^2(l+1)^2\over
k^4}\delta_{lp}\delta_{mq} , \label{q1} \end{equation} and
\begin{equation} Q^{(2)}_{lmpq}(k)\equiv\int d\Omega_{\hat k}
\big({\tilde\nabla}_a{\tilde\nabla}_b Y_{pq}\big)\big({\tilde\nabla}^a
{\tilde\nabla}^b Y^*_{lm}\big).  \label{q22} \end{equation} Integrating
(\ref{q22}) by parts once, we find \begin{equation}
Q^{(2)}_{lmpq}(k)=-\int d\Omega_{\hat k} \big({\tilde\nabla}_b
Y_{pq}\big)\big( {\tilde{\smash{\Box}\vphantom{\nabla}}}
{\tilde\nabla}^b Y^*_{lm}\big), \end{equation} since the two-sphere has
no boundary. With the identity (\ref{yident}) we have \begin{equation}
Q^{(2)}_{lmpq}(k)=-\bigg[{-l(l+1)+1\over k^2}\bigg] \int d\Omega_{\hat
k} \big({\tilde\nabla}_b Y_{pq}\big)\big( {\tilde\nabla}^b
Y^*_{lm}\big).  \end{equation} One may again integrate by parts and use
the eigenfunction equation for the spherical harmonics to obtain
\begin{eqnarray} Q^{(2)}_{lmpq}(k)&=&\bigg[{-l(l+1)+1\over k^2}\bigg]
\bigg[{-l(l+1)\over k^2}\bigg]\delta_{lp}\delta_{mq},\\ &=&{1\over
k^4}\big[l^2(l+1)^2-l(l+1)\big] \delta_{lp}\delta_{mq} .  \label{q2}
\end{eqnarray} Substituting (\ref{q1}) and (\ref{q2}) into
(\ref{awithq}) one has \begin{equation} A_{lmpq}(k,r,r')= {16\pi^2
\over r^2 r'^2 k^4}j_l(kr')j_l(kr)
\>(l-1)l(l+1)(l+2)\>\delta_{lp}\delta_{mq} .  \label{afinal}
\end{equation} As previously indicated, the matrix element ${\langle
0|{\bar C}_{pq}^\dagger \bar C_{lm}|0\rangle}$ and the correlation
function $C(\hat v^c,\hat u^c)$ are indeed rotationally invariant.
Also note that (\ref{afinal}) vanishes for $l=0$ and $l=1$.

\subsubsection{The Angular Correlation Function}

Using the above form of $A_{lmpq}(k,r,r')$ one may derive a simple
expression for the angular correlation function $C(\hat v^c,\hat u^c)$,
and show directly that it depends only on the angle $\gamma$ between
$\hat v^c$ and $\hat u^c$.  With (\ref{afinal}) and (\ref{cwitha}) one
has \begin{eqnarray} {\langle 0|{\bar C}_{pq}^\dagger \bar
C_{lm}|0\rangle}=&& 4\pi^2{(l+2)!\over
(l-2)!}\delta_{lp}\delta_{mq}\nonumber\\ &&\times\int\limits^\infty_0
{dk\over k} \int\limits^{{\lambda_{\rm obs}} }_{{\lambda_{\rm e}}
}d\lambda ' \int\limits^ {{\lambda_{\rm obs}} }_{{\lambda_{\rm e}}
}d\lambda \>F(\lambda
',k)F^{*}(\lambda,k)\>{j_l(kD(\lambda))j_l(kD(\lambda')) \over k^2
D^2(\lambda) D^2(\lambda')}, \end{eqnarray} where we have written the
fourth-order polynomial in $l$ appearing in (\ref{afinal}) as the ratio
of two factorials.  Noting the symmetry of the right-hand side, and
recalling the definitions of $D(\lambda)$, ${\lambda_{\rm e}} $, and
${\lambda_{\rm obs}} $, we define
 \begin{equation} I_l(k)\equiv\int^{{\eta_{\rm obs}}-{\eta_{\rm e}}} _0
d\lambda\>F(\lambda,k) {j_l(k(\eta_{obs}-\eta_e-\lambda))\over k
(\eta_{obs}-\eta_e-\lambda)^2} , \label{ilk} \end{equation} and write
the matrix element as
 \begin{equation} {\langle 0|{\bar C}_{pq}^\dagger \bar
C_{lm}|0\rangle}= \langle a_l^2\rangle \delta_{lp}\delta_{mq},
\end{equation} where \begin{equation} \langle a_l^2\rangle\equiv
4\pi^2{(l+2)!\over (l-2)!} \int^\infty_0 {dk\over k} \>\>|I_l(k)|^2.
\label{aldefinition} \end{equation} Substituting this expression into
(\ref{cexpan}) we obtain \begin{equation} C(\hat v^c,\hat u^c)
=\sum_{lm}\langle a_l^2\rangle Y_{lm}(\hat u^c) Y^*_{lm}(\hat v^c),
\end{equation} where we have used the Kronecker deltas to eliminate two
of the sums.  Making use of the addition theorem for spherical
harmonics (see equation 3.62 in reference \cite{Jackson}), the
correlation function is \begin{equation} C(\hat v^c,\hat u^c)\equiv
C(\gamma)=\sum_{l=0}^\infty {(2l+1)\over 4\pi} \langle a_l^2\rangle
P_l(\cos\gamma), \label{cofgamma} \end{equation} where \begin{equation}
\cos\gamma\equiv\hat u^c \hat v_c .  \end{equation} As promised, the
angular correlation function depends only on the angle $\gamma$ between
any two points on the celestial sphere. Also note that the $l=0$ and
the $l=1$ terms in the expansion vanish exactly.

This form of the correlation function is very general.  The only
dependence of the correlation function on the details of any
cosmological model is through the graviton mode function (or more
precisely, its first derivative), which appears as $F(\lambda,k)$ in
the definition of $I_l(k)$ (\ref{ilk}). A similar result, which is as
general as (\ref{cofgamma}), is given by Grishchuk (see equation 4 in
the second paper of reference \cite{Grishchuk2}).


\section{Graviton Mode Function Normalization}
\label{section3}

If one demands that the metric perturbation field
operator $\bar h_{ab}$ obey canonical commutation relations,
the commutation relations (\ref{acommutation}) for the graviton
creation and annihilation operators imply that the
graviton mode function satisfy a normalization condition.
Imposing canonical commutation relations on the {\it tensor}
field $\bar h_{ab}$, however, is subtle because
as noted by Ford and Parker \cite{FordParker},
the canonical commutation relation  that $\bar h_{ab}$ obeys may be
inconsistent with the gauge conditions
on $\bar h_{ab}$. For this reason, we
follow reference \cite{FordParker} and impose canonical
commutation relations on the two independent
{\it scalar} degrees of freedom in $\bar h_{ab}$.

The two independent degrees of freedom in the metric perturbation field
can be isolated by constructing two scalar field operators from $\bar
h_{ab}$.  Recall from (\ref{barhab}) that the plane wave expansion of
$\bar h_{ab}$ is \begin{equation} \bar h_{ab}(\eta,x^c)=\int d^3k\>\bar
h_{ab}(\eta,x^c,k^c), \end{equation} where \begin{eqnarray} \bar
h_{ab}(\eta,x^c,k^c)= \bigg(& &e^{i  k^d x_d}\big[e_{ab}(
k^c)\phi_R(\eta,k^c) \bar a_R( k^c) +e^*_{ab}(k^c)\phi_L(\eta,k^c)\bar
a_L( k^c) \big]\nonumber\\
 &&+ e^{-i  k^d x_d}\big[e^*_{ab}(k^c) \phi_R^*(\eta,k^c) \bar
a^\dagger_R( k^c) +e_{ab}(k^c) \phi_L^*(\eta,k^c)\bar a^\dagger_L( k^c)
\big]\bigg).  \end{eqnarray} We define the scalar field operator
\begin{equation} \bar h_+(\eta,x^c)\equiv\int d^3k\>\bar
h_{ab}(\eta,x^c,k^c) e^{(+)}{}^{ab}(k^c).  \label{hplus1}
\end{equation} Contracting the integrand using
(\ref{orthog}-\ref{circpolar}) one obtains \begin{eqnarray} \bar
h_+(\eta,x^c)&=&\sqrt{2}\int d^3k\> \bigg\{e^{i  k^d x_d}\phi(\eta,k)
\big[\bar a_R( k^c) +\bar a_L( k^c) \big]\nonumber\\ && + e^{-i  k^d
x_d}\phi^*(\eta,k)
 \big[ \bar a_R( k^c) +\bar a_L( k^c) \big]^\dagger\bigg\}.
\label{hplus2} \end{eqnarray} A second scalar field operator $\bar
h_\times(\eta,x^c)$ is defined by replacing the plus signs $(+)$ in
(\ref{hplus1}) by crosses $(\times)$, which has the effect of replacing
$\bar a_R+\bar a_L$ by $i\bar a_R-i\bar a_L$ in (\ref{hplus2}).
Together the scalar field operators $\bar h_+(\eta,x^c)$ and $\bar
h_\times(\eta,x^c)$ possess the same two degrees of freedom as the
metric perturbation operator $\bar h_{ab}$ \cite{FordParker}.  Since
$\bar h_+ (\eta,x^c)$ and $\bar h_\times(\eta,x^c)$ are both {\it
scalar} field operators, they obey  well known canonical commutation
relations for scalar fields. Because our particle basis does not
distinguish between the two polarizations, we only need to consider one
of the two scalar fields, since both lead to the same normalization
condition for the mode function.

The scalar field operator $\bar h_+(\eta,x^c)$ obeys the canonical
commutation relation \begin{equation} \big[\bar
h_+(\eta,x^c),\bar\pi_+(\eta,x'^c) \big]=i\hbar \delta^3(x^c-x'^c),
\label{hpicommutation} \end{equation} where the field operator
$\bar\pi_+(\eta,x^c)$ is canonically conjugate to $\bar h_+
(\eta,x^c)$, and is defined by \begin{equation}
\bar\pi_+\equiv{\delta{\cal L}\over\delta \dot{\bar h}_+}.
\label{pidefinition} \end{equation} Here $\cal{L}$ is the Lagrangian
density of the perturbed spatially-flat FRW spacetime.  To impose the
commutation relation (\ref{hpicommutation}) and find the normalization
condition for the graviton mode function we need to find the Lagrangian
density in terms of $\bar h_+$ and $\bar h_\times$.

The Lagrangian density is obtained by expanding the gravitational plus
matter action to second order in the metric perturbation $\bar
h_{ab}$.  The action for a FRW spacetime is \begin{equation} S=\int
d^4x \sqrt{g}\bigg\{{R\over 16\pi G}+ {1\over 2}[(\rho+P)u^\mu u^\nu
g_{\mu\nu}+(\rho+3P)] \bigg\}, \label{action} \end{equation} where
$g_{\mu\nu}$ is the metric for the FRW spacetime, $R$ is the Ricci
scalar, $\rho$ is the energy density, $P$ is the pressure , and $u^\mu$
is the four velocity of the cosmological fluid.  Varying the action
with respect to $g_{\mu\nu}$ leads to the Einstein equation for a FRW
model:  \begin{equation} G_{\mu\nu}=R_{\mu\nu}+{1\over 2} R g_{\mu\nu}
=8\pi G T_{\mu\nu}=8\pi G[(\rho+P)u_\mu u_\nu+Pg_{\mu\nu}].
\label{einsteinequation} \end{equation} If the FRW spacetime is
perturbed so that \begin{equation}
g_{\mu\nu}={^0g}_{\mu\nu}+\gamma_{\mu\nu}, \end{equation} where
${^0g}_{\mu\nu}$ is the unperturbed or background FRW metric, but the
pressure $P$ and the energy density $\rho$ are {\it not} perturbed,
then one finds that the second-order variation in the action
is\cite{FordParker} \begin{eqnarray} \delta^2 S=\int d^4 x
&&\sqrt{^0g}{1\over 64\pi G}\bigg\{
(^0\nabla^\mu\gamma^{\nu\xi})(^0\nabla_\mu\gamma_{\nu\xi})+ 8\pi
G(P-\rho)\gamma^{\mu\nu}\gamma_{\mu\nu}\nonumber\\ &&+2
{}^0\negthinspace R_{\mu\nu} \gamma^{\mu\xi}\gamma_\xi{}^\nu+2
{}^0\negthinspace R_{\mu\nu\xi\sigma}
\gamma^{\mu\sigma}\gamma^{\nu\xi}\bigg\}.  \label{deltaS}
\end{eqnarray} The superscript, for example in $^0\nabla_\mu$, refers
to the background spacetime. One should note that (\ref{deltaS}) is
obtained by making a specific choice of gauge (transverse,
traceless)\cite{FordParker}.  Also note that this is to {\it second}
order in the perturbation $\gamma_{\mu\nu}$, since the first order part
$\delta S$ vanishes because the background FRW spacetime satisfies
(\ref{einsteinequation}).

Equation (\ref{deltaS}) is very general and true for any ``small''
perturbation $\gamma_{\mu\nu}$ (that satisfies the gauge conditions)
away from a FRW spacetime with metric ${^0g}_{\mu\nu}$.  For our
purposes, the perturbation $\gamma_{\mu\nu}$ is simply
$a^2(\eta)h_{ab}$, and the background FRW spacetime is spatially-flat.
With a little calculation one can show that for the spatially-flat FRW
spacetime perturbed by tensor perturbations \begin{equation} \delta^2
S=\int d^4x {a^2(\eta)\over 64\pi G}\bigg\{ -{\dot h}_{ab}{\dot
h}^{ab}+(\partial_a h_{bc}) (\partial^a h^{bc})\bigg\}.  \end{equation}
To calculate the momentum $\bar\pi_+$ conjugate to $\bar h_+$, one must
express the action in terms of $\bar h_+$ and $\bar h_\times$. With a
little calculation, and using (\ref{barhab}), one can write the action
as \begin{equation} \delta^2 S=\int d^4 x {a^2(\eta)\over 64\pi
G}{1\over 2} \{-(\dot{\bar h}_+^2+\dot{\bar h}_\times^2)
+(\partial_a\bar h_+)(\partial^a\bar h_+)+ (\partial_a\bar
h_\times)(\partial^a\bar h_\times)\}, \end{equation} so that the
Lagrangian density is \begin{equation} {\cal L}={a^2(\eta)\over 64\pi
G}{1\over 2} \{-(\dot{\bar h}_+^2+\dot{\bar h}_\times^2)
+(\partial_a\bar h_+)(\partial^a\bar h_+)+ (\partial_a\bar
h_\times)(\partial^a\bar h_\times)\}.  \label{lagrangian}
\end{equation} This is just the Lagrangian for a pair of massless
scalar fields minimally coupled to the background spacetime.  Using
(\ref{pidefinition}) and (\ref{lagrangian}) the momentum is
\begin{equation} \bar\pi_+(\eta,x^a)=-{a^2(\eta)\over 64\pi G }
\dot{\bar h}_+ .  \end{equation} Then from (\ref{hpicommutation}) the
canonical commutation relation for the scalar field operator $\bar h_+$
is \begin{equation} \big[\bar h_+(\eta,x^a),\dot{\bar h}_+(\eta,x'^a)
\big]=-64\pi i\hbar\> G\> {\delta^3(x^a-x'^a)\over a^2(\eta)}.
\label{hcommutation} \end{equation} Note that this is an equal-time
commutation relation.

Using the commutation relation above and the explicit form  for the
scalar field $\bar h_+$, one can derive the normalization condition for
the graviton mode function. Using (\ref{hplus2}) and
(\ref{acommutation}) one finds \begin{equation} \big[\bar
h_+(\eta,x^a),\dot{\bar h}_+(\eta,x'^a) \big]=4\int d^3k
\big\{\phi(\eta,k)\dot{\phi^*}(\eta,k) e^{i
k^a(x_a-x'_a)}-\phi^*(\eta,k)\dot\phi(\eta,k) e^{-i
k^a(x_a-x'_a)}\big\}.  \end{equation} Since we assume that the mode
function $\phi(\eta,k)$ depends only on the magnitude $k$, one can
write \begin{equation} \big[\bar h_+(\eta,x^a),\dot{\bar
h}_+(\eta,x'^a) \big]=4\int d^3k e^{i k^a(x_a-x'_a)}
\big\{\phi(\eta,k)\dot{\phi^*}(\eta,k)
-\phi^*(\eta,k)\dot\phi(\eta,k)\big\}.  \label{almostwronskian}
\end{equation} Since the delta function in (\ref{hcommutation}) can be
expressed as a plane-wave expansion \begin{equation} \int d^3k e^{i
k^a(x_a-x'_a)}= (2\pi)^3\delta^3(x^c-x'^c), \end{equation}
(\ref{hcommutation}) and (\ref{almostwronskian}) imply the mode
function normalization condition \begin{equation}
\big\{\phi(\eta,k)\dot{\phi^*}(\eta,k)
-\phi^*(\eta,k)\dot\phi(\eta,k)\big\} =-{2 i\hbar G\over \pi^2
a^2(\eta)}.  \label{phinormalization} \end{equation} This identity
determines the normalization of the graviton mode function, up to an
(irrelevant) overall phase,  and would be equivalent to equation (3.3)
of reference \cite{FordParker} if not for a typo  \cite{Ford2}.  The
main consequence is that fundamental physical principles (the
uncertainty principle) completely determine the amplitude of the
contribution to the angular correlation function arising from
gravitational radiation.

\section{INFLATIONARY COSMOLOGICAL MODEL}
\label{section4}

\subsection{Graviton Mode Function}
\label{section4b}

The cosmological model we examine ``begins'' with an infinite
inflationary phase, followed by radiation- and then  matter-dominated
phases.  We assume that the inflationary phase evolves into de Sitter
spacetime. The mechanism by which the universe arrives at the de Sitter
spacetime is not important, since the period of rapid expansion during
the de Sitter phase effectively erases the initial conditions.  At the
end of the  de Sitter phase the universe undergoes an instantaneous
phase transition to a radiation-dominated FRW phase. At the end of the
radiation phase the universe again undergoes an instantaneous phase
transition, and evolves as a  matter-dominated FRW spacetime until the
present.

If the initial de Sitter phase is sufficiently long, the spatial
geometry becomes flat, and one may assume that the universe is
spatially-flat for all three epochs. The metric for the spacetime is
then given by (\ref{ourmetric}), with scale factor \begin{equation}
a(\eta)=\left\{\begin{array}{lll} \big(2-{\eta\over{\eta_1}}\big)^{-1}
a({\eta_1}) & -\infty<\eta<{\eta_1} &{\rm de Sitter,}\\ &&\\
{\eta\over{\eta_1}} a({\eta_1}) & {\eta_1}<\eta<{\eta_2}  & {\rm
radiation,}\\ &&\\ {1\over 4}\big(1+{\eta\over{\eta_2} }\big)^2
{{\eta_2} \over{\eta_1}} a({\eta_1}) & {\eta_2} <\eta &{\rm matter},
\end{array} \right.  \label{scalefactor} \end{equation} where
${\eta_1}$ and ${\eta_2} $ are constants. The redshift at the end of
the de Sitter phase $Z_{\rm end}$ and the redshift at the time of
radiation--matter equality $Z_{\rm equal}$ are defined by
\begin{eqnarray} 1+Z_{\rm end}&&\equiv{a({\eta_{\rm obs}})\over
a({\eta_1})}={({\eta_{\rm obs}}+{\eta_2} )^2\over 4{\eta_1}{\eta_2}
},\label{Zenddefinition}\\ 1+Z_{\rm equal}&&\equiv{a({\eta_{\rm
obs}})\over a({\eta_2} )}={({\eta_{\rm obs}}+{\eta_2} )^2\over
4{\eta_2} ^2}, \label{Zequaldefinition} \end{eqnarray} where
${\eta_{\rm obs}}$ is conformal time today.  Typical values for the
redshifts (for models ``with enough inflation'' to solve the horizon
and flatness problems) are $Z_{\rm end}\approx 10^{27}$ and $Z_{\rm
equal}\approx 10^4$.  We assume that  last-scattering  at conformal
time ${\eta_{\rm e}}$ took place after the time of radiation--matter
equality so that  $\eta_e>{\eta_2} $. The reshift of the surface of
last-scattering $Z_{\rm ls}$ is \begin{equation} 1+Z_{\rm
ls}\equiv{a({\eta_{\rm obs}})\over a({\eta_{\rm e}})}=\bigg({{\eta_{\rm
obs}}+{\eta_2} \over {\eta_{\rm e}}+{\eta_2} }\bigg)^2.
\label{Zlsdefinition} \end{equation} A typical value for the redshift
of the surface of last-scattering is $Z_{\rm ls}\approx 1300$, although
it is possible that the hydrogen  was re-ionized as recently as
redshift $Z_{\rm ls}\approx 100$ \cite{JonesWyse}.

Note that the scale factor (\ref{scalefactor}) and its first derivative
are continuous. Because the second derivative of the scale factor is
not continuous, the scalar curvature of the spacetime changes
discontinuously at the phase transitions. This instantaneous  phase
transition is a good approximation, except at high frequencies, where
it predicts too much graviton production \cite{Allen}.

With the scale factor above, one can solve the massless Klein-Gordon
equation (\ref{KleinGordon}) for the graviton mode function during each
of the three epochs.  By making a change of dependent, and then
independent variable, the Klein-Gordon equation can be cast in the form
of Bessel's  equation, for each of the three phases. The necessary
changes of variable, and the positive-frequency solutions, are shown in
table \ref{table1}. Using the normalization condition
(\ref{phinormalization}), and making a convenient choice of phase, one
obtains the following positive-frequency solutions for the three
epochs:  \begin{eqnarray} \phi^{(+)}_{\rm ds}(\eta,k)&=
&-i\sqrt{{8\over 3\pi} {\rho_{\rm ds}\over\rho_{\rm p}}}\>k^{1/2}
\>(\eta-2{\eta_1})^2\>h_1^{(2)}(k(\eta-2{\eta_1})) e^{-ik{\eta_1}},
\>\text{for the de Sitter phase,}\label{deSittermodefunction}\\
&&\nonumber\\ \phi^{(+)}_{\rm rad}(\eta,k)&=&-i \sqrt{{8\over 3\pi}
{\rho_{\rm ds}\over\rho_{\rm p}}}\>k^{1/2}
\>{\eta_1}^2\>h_0^{(2)}(k\eta) e^{ik{\eta_1}},\> \text{for the
radiation phase,}\\ &&\nonumber\\ \phi^{(+)}_{\rm mat}(\eta,k)&=
&-4i\sqrt{{8\over 3\pi} {\rho_{\rm ds}\over\rho_{\rm p}}}\>k^{1/2}
\>{\eta_1}^2{\eta_2} \>{h_1^{(2)}(k(\eta+{\eta_2} ))\over \eta+{\eta_2}
},\>\text{for the matter phase}, \label{mattermodefunction}
\end{eqnarray} where \begin{equation} \rho_{\rm ds}={3\over 8\pi G}
{{\dot a}^2({\eta_1})\over a^4({\eta_1})}= {3\over 8\pi G}{1\over
{\eta_1}^2 a^2({\eta_1})} \end{equation} is the (constant) energy
density during the de Sitter phase, and \begin{equation} \rho_{\rm
p}={1\over \hbar G^2} \end{equation} is the Planck energy density. The
spherical Hankel functions \cite{Jackson} are defined by
\begin{equation} h^{({1\atop 2})}_l(z)=j_l(z)\pm i y_l(z),
\label{hankelfunction} \end{equation} where $j_l(z)$ and $y_l(z)$ are
spherical Bessel functions of the first and second kind.  The
negative-frequency mode functions are the complex conjugates of the
positive-frequency mode functions.  The positive- and
negative-frequency solutions for each epoch form a complete set of
solutions to the massless Klein-Gordon equation (\ref{KleinGordon}).

The choice of a mode function during the initial de Sitter phase
$\eta<\eta_1$ completely determines the mode function at all later
times. This is because a solution to the Klein-Gordon equation
(\ref{KleinGordon}) depends only upon the values of $\phi$ and
$\dot\phi$ on a spacelike hypersurface (i.e. a surface of fixed
$\eta$). To express the solution $\phi$ at later times, after the de
Sitter phase has ended, it is useful to adopt the {\it Bogolubov
coefficient} notation. In this notation, the solution $\phi$ at later
times is expressed as a linear combination
$\alpha\phi^{(+)}+\beta\phi^{(-)}$ of the natural choices of positive
and negative frequency solutions $\phi$ during the subsequent phases of
expansion.

If one evolves the positive-frequency mode function during the de
Sitter phase $\phi^{(+)}_{\rm ds}$ into the subsequent radiation phase
via (\ref{KleinGordon}), it is necessary that the  mode function and
its first derivative be continuous across the phase transitions at
$\eta={\eta_1}$ and $\eta={\eta_2} $.  Continuity from the de Sitter to
the radiation phase is assured if and only if the Bogolubov
coefficients $\alpha_{\rm rad}$ and $\beta_{\rm rad}$ satisfy the
conditions \begin{eqnarray} \phi^{(+)}_{\rm
ds}({\eta_1},k)&=&\alpha_{\rm rad} \phi^{(+)}_{\rm rad}({\eta_1},k)+
\beta_{\rm rad}\phi^{(-)}_{\rm rad}({\eta_1},k),\nonumber\\
\dot\phi^{(+)}_{\rm ds}({\eta_1},k)&=&\alpha_{\rm rad}
\dot\phi^{(+)}_{\rm rad}({\eta_1},k)+ \beta_{\rm
rad}\dot\phi^{(-)}_{\rm rad}({\eta_1},k).  \end{eqnarray} Solving this
pair of linear equations one finds \begin{eqnarray} \alpha_{\rm
rad}({\eta_1},k)&=&-i\bigg(1+{i\over k{\eta_1}} -{1\over 2
k^2{\eta_1}^2}\bigg),\nonumber\\ \beta_{\rm rad}({\eta_1},k)&=&{i\over
2 k^2 {\eta_1}^2}.  \end{eqnarray} Likewise, if one evolves the
positive-frequency mode function during the radiation phase
$\phi^{(+)}_{\rm rad}$ into the subsequent matter phase, the Bogolubov
coefficients $\alpha_{\rm mat}$ and $\beta_{\rm mat}$ must satisfy
\begin{eqnarray} \phi^{(+)}_{\rm rad}({\eta_2} ,k)&=&\alpha_{\rm mat}
\phi^{(+)}_{\rm mat}({\eta_2} ,k)+ \beta_{\rm mat}\phi^{(-)}_{\rm
mat}({\eta_2} ,k),\nonumber\\ \dot\phi^{(+)}_{\rm rad}({\eta_2}
,k)&=&\alpha_{\rm mat} \dot\phi^{(+)}_{\rm mat}({\eta_2} ,k)+
\beta_{\rm mat}\dot\phi^{(-)}_{\rm mat}({\eta_2} ,k).  \end{eqnarray}
Solving this pair of linear equations we find \begin{eqnarray}
\alpha_{\rm mat}({\eta_2} ,k)&=&-i\bigg(1+{i\over 2k{\eta_2} } -{1\over
8 k^2{\eta_2} ^2}\bigg)e^{ik({\eta_1}+{\eta_2} )},\nonumber\\
\beta_{\rm mat}({\eta_2} ,k)&=&{i\over 8 k^2 {\eta_2} ^2}
e^{ik({\eta_1}-3{\eta_2} )}.  \label{mattercoefficients} \end{eqnarray}
Since the mode functions are normalized by (\ref{phinormalization}),
the Bogolubov coefficients obey the (easily verified) relation
\begin{equation} |\alpha_{\rm rad}|^2-|\beta_{\rm rad}|^2 =
|\alpha_{\rm mat}|^2-|\beta_{\rm mat}|^2 =1.  \end{equation} The
Bogolubov coefficients above agree with reference \cite{Allen}, up to
an irrelevant phase.

As stated earlier, the choice of a mode function during the de Sitter
phase completely determines the mode function at all later times. We
choose the mode function for the de Sitter phase to be the
positive-frequency de Sitter solution (\ref{deSittermodefunction}):
\begin{equation} \phi(\eta,k)=\phi^{(+)}_{\rm ds}(\eta,k),
\>\>\text{for}\>\>  -\infty<\eta<{\eta_1} .  \label{positivefreq}
\end{equation} This is the unique solution corresponding to a de
Sitter-invariant vacuum state with the same (Hadamard) short distance
behavior as one would find in Minkowski space \cite{Allen}.  Having
calculated the Bogolubov coefficients, one may now determine the way in
which the positive-frequency mode function (\ref{positivefreq}) evolves
continuously from one phase to the next.  The complete mode function
during all three epochs is \begin{equation}
\phi(\eta,k)=\left\{\begin{array}{llll} \phi^{(+)}_{\rm ds}(\eta,k)
&\text{for}& -\infty<\eta<{\eta_1} & {\rm de Sitter},\\ &&&\\
\alpha_{\rm rad} \phi^{(+)}_{\rm rad}(\eta,k)+ \beta_{\rm
rad}\phi^{(-)}_{\rm rad}(\eta,k) &\text{for}& {\eta_1}<\eta<{\eta_2}  &
{\rm radiation,}\\ &&&\\ \alpha\>\phi^{(+)}_{\rm mat}(\eta,k)
+\beta\>\phi^{(-)}_{\rm mat}(\eta,k) &\text{for}& {\eta_2} <\eta & {\rm
matter}, \end{array} \right.  \label{finalmodefunction} \end{equation}
where the coefficients $\alpha$ and $\beta$ are given by
\begin{equation} \pmatrix{\alpha &\beta\cr \beta^* &\alpha^*}=
\pmatrix{\alpha &\beta\cr \beta^* &\alpha^*}_{\rm rad} \pmatrix{\alpha
&\beta\cr \beta^* &\alpha^*}_{\rm mat}.  \label{alphabetadefinition}
\end{equation} The mode function (\ref{finalmodefunction}) is the
normalized, continuous, graviton mode function which appears in the
expression for the correlation function (\ref{ilk}).  This expression
for the mode function is {\it exact}, and valid for all wavenumbers
$k$.

\subsection{Corrections to the Instantaneous Phase \protect\\
Transition Approximation}

Our inflationary cosmological model  undergoes {\it instantaneous}
phase transitions; first between the de Sitter and radiation phase, and
then between the radiation and matter phase.  At these transitions, the
scalar curvature of the universe changes abruptly, since the second
derivative of the scale factor (\ref{scalefactor}) is discontinuous.
This abrupt change in curvature produces gravitons, much in the same
way as an abrupt change in the electromagnetic potential produces
photons.  This instantaneous phase transition is a good approximation,
except at high frequencies, where it predicts too much graviton
production \cite{Allen}.

The physical universe transforms smoothly from phase to phase, with
each transition taking place during a characteristic period of time.
If the characteristic time of a phase transition is $\Delta t$, then
one would expect the spectrum of gravitons produced by the phase
transition to be supressed above a cut-off frequency $f_{\rm cut}$,
with $f_{\rm cut}\sim 1/\Delta t$. Equivalently, the production of
gravitons whose wavelength is less than $\lambda_{\rm cut}=\Delta t$ is
supressed.  The idealization that the phase transitions are
instantaneous is a good approximation for frequencies below $f_{\rm
cut}$. For this reason, the multipole moments $\langle a_l^2\rangle$
for small values of $l$ should be unaffected by this idealization.  The
$\langle a_l^2\rangle$ for large $l$, however, will be overestimated if
we do not ``smooth out'' the phase transitions.

The adiabatic theorem \cite{ParkerCreation,BirrellDavies}  provides a
simple way to account for the effects of ``smoothing out'' the phase
transition, which does not require any detailed information about how
the abrupt change in $\ddot{a}(\eta)$ is smoothed.  The cut-off
wavelength $\lambda_{\rm cut}$ corresponds to a cut-off wavenumber
$k_{\rm cut}$. The adiabatic theorem implies that the Bogolubov
coefficient $\beta$ in (\ref{finalmodefunction}), whose modulus squared
gives the number of gravitons produced in a given mode \cite{Allen},
should decay exponentially for $k>k_{\rm cut}$, while the Bogolubov
coefficient $\alpha$ goes exponentially to 1.  So for large values of
$k$ with $k\gg k_{\rm cut}$ no graviton production takes place.

To be specific, consider the transition from the radiation  to the
matter phase.  We will assume that the characteristic time of the phase
transition equals the Hubble length at that time.  This gives a cut-off
wavelength \begin{equation} \lambda_{\rm cut}({\eta_2} )={a^2({\eta_2}
)\over {\dot a}({\eta_2} )}={\rm H}^{-1}({\eta_2} ).  \end{equation} At
later times, the cut-off wavelength is redshifted by the cosmological
expansion to the longer wavelength \begin{equation} \lambda_{\rm
cut}(\eta)=\lambda_{\rm cut}({\eta_2} ){a(\eta) \over a({\eta_2} )}.
\end{equation} The cut-off (comoving) wavenumber is then given by
\begin{equation} k_{\rm cut}={2\pi a(\eta)\over\lambda_{\rm cut}(\eta)}
=2\pi{\eta_2} ^{-1}.  \end{equation} The adiabatic theorem now implies
that for comoving wavenumbers above $k_{\rm cut}$ \begin{eqnarray}
\alpha_{\rm mat}(k>k_{\rm cut})&=&\bigg[ {1+|\beta_{\rm mat}(k_{\rm
cut})|^2 e^{2-2k/k_{\rm cut}}\over 1+|\beta_{\rm mat}(k_{\rm
cut})|^2}\bigg]^{1/2} \alpha_{\rm mat} (k_{\rm cut}),\\ &&\nonumber\\
\beta_{\rm mat}(k>k_{\rm cut})&=&\beta_{\rm mat}(k_{\rm cut})
e^{1-k/k_{\rm cut}}.  \label{betterbogoliubov} \end{eqnarray} We use
these formulae to determine $\alpha_{\rm mat}$ and $\beta_{\rm mat}$
for $k>k_{\rm cut}$. They only significantly effect multipole moments
$\langle a_l^2\rangle$ with $l\gtrsim 1000$. A similar analysis shows
that the instantaneous transition from the de Sitter to the radiation
phase  only affects the moments with extremely large $l$.

\subsection{The Long Wavelength Approximation}

As noted in the introduction, the previously published calculations
determine the angular correlation function using a ``long wavelength''
approximation to the graviton mode function. (We assume that the
last-scattering event took place after the universe became
matter-dominated, ie.  ${\eta_2} <{\eta_{\rm e}}\>$; for the rest of
this paper, the ``mode function'' means the mode function during the
{\it matter} phase).  The long wavelength approximation is the same as
an approximation for small wavenumber $k$. To make a small $k$
approximation to the mode function, it is helpful to express the mode
function in terms of  spherical Bessel functions. Using
(\ref{mattermodefunction}), (\ref{hankelfunction}), and
(\ref{finalmodefunction}), one can write the mode function as
\begin{equation} \phi(\eta,k)=-4i\sqrt{{8\over 3\pi}{\rho_{\rm ds}\over
\rho_{\rm p}}} {k^{1\over 2} {\eta_1}^2{\eta_2}  \over(\eta+{\eta_2} )}
\bigg\{[\alpha +\beta ]j_1(k(\eta+{\eta_2} )) -i[\alpha -\beta
]y_1(k(\eta+{\eta_2} )) \bigg\}.  \label{plusandminus} \end{equation}
To understand the small $k$ behavior of $\phi(\eta,k)$, one can expand
the combinations of Bogolubov coefficients in square brackets as power
series in $k$. One finds \begin{eqnarray} [\alpha +\beta ]=&&{3i\over
4}{1\over{\eta_1}^2 {\eta_2}  k^3}+{\rm O}(k^{-2})\nonumber\\ \protect
[\alpha - \beta ]=&&\bigg({40{\eta_1}^3{\eta_2} -4{\eta_2}
^4\over45{\eta_1}^2}\bigg)k^2+ {\rm O}(k^3).  \label{smallkbogoliubov}
\end{eqnarray} Furthermore, the small $k$ behavior of the spherical
Bessel functions can be understood by noting that for small argument,
\begin{equation} z\ll 1\>\Rightarrow\>j_1(z)\sim {z\over 3}
\>\>\text{and}\>\>y_1(z)\sim -z^{-2}.  \label{smallbessel}
\end{equation} Using (\ref{smallkbogoliubov}) and (\ref{smallbessel}),
and noting that \begin{equation} j_l(z)=\sqrt{\pi\over 2 z}
J_{l+1/2}(z), \label{besselfunctionrelation} \end{equation} one sees
that for small wavenumber $k$ one may approximate the graviton mode
function by \begin{equation} \phi(\eta,k)\approx\sqrt{12{\rho_{\rm
ds}\over \rho_{\rm p}}}\> {J_{3/2}(k(\eta+{\eta_2} ))\over k^3
(\eta+{\eta_2} )^{3/2}}\>\>\text{for}\>\>{\eta_2}
<\eta\>\>\text{(matter)}.  \label{phiapprox} \end{equation} The
validity of this small $k$ or long wavelength approximation, in the
context of the angular correlation function, is discussed in section
\ref{section6}.

Using (\ref{plusandminus}) and (\ref{smallkbogoliubov}) one can see why
the long wavelength tensor perturbations can be thought of as classical
gravitational waves.  The Bogolubov coefficients are restricted by the
constraint $|\alpha |^2 - |\beta|^2 = 1$, and in realistic inflationary
models the ``occupation number" $|\beta|^2$ \cite{BirrellDavies} is
much greater than one for small wavenumber $k$. Hence for small
wavenumber $\alpha$ and $\beta$ are both very large, and almost equal,
as is apparent from (\ref{smallkbogoliubov}). In this limit the graviton
mode function (\ref{phiapprox}) may be thought of and treated as a
classical gravitational wave, as one would expect for a bosonic field
with large occupation number.

\subsection{Standard Results}
\label{section4d}

Using the long wavelength approximation to the graviton mode function
we can reproduce the standard results
\cite{Rubakov,Fabbri,AbbotWise,Starobinsky2,AbbottShaefer} for the
angular correlation function due to gravitational-wave perturbations.
Although the long wavelength approximation is valid only for small
wavenumber $k$, we assume it to hold for all $k$.  Recall that the {\it
first derivative} of the mode function (\ref{Fdefinition}) appears in
the angular correlation function. With the approximate mode function
(\ref{phiapprox}) and the definition (\ref{Fdefinition}) for the
function $F(\lambda,k)$, one finds \begin{equation}
F(\lambda,k)=\sqrt{12{\rho_{\rm ds}\over \rho_{\rm p}}}\>
{J_{5/2}(k({\eta_2} +{\eta_{\rm e}}+\lambda))\over k^{3/2} ({\eta_2}
+{\eta_{\rm e}}+\lambda)^{3/2}}.  \label{Fapproximation} \end{equation}
Note that we have used the standard recurrence relations (equation
(9.1.27) of reference \cite{AbramowitzStegun}) for Bessel functions to
put $F(\lambda,k)$ is this form.  Substituting this into (\ref{ilk})
and using (\ref{besselfunctionrelation}) one obtains \begin{equation}
I_l(k)=\sqrt{{6\pi}{\rho_{\rm ds}\over \rho_{\rm
p}}}\>k^{-3}\>\int\limits_0^{{\eta_{\rm obs}}-{\eta_{\rm e}}}d\lambda
{J_{5/2}(k({\eta_2} +{\eta_{\rm e}}+\lambda))J_{l+1/2}(k({\eta_{\rm
obs}}-{\eta_{\rm e}}-\lambda)) \over ({\eta_2} +{\eta_{\rm
e}}+\lambda)^{3/2}  ({\eta_{\rm obs}}-{\eta_{\rm e}}-\lambda)^{5/2}}.
\end{equation} To compare our results with  standard formulae, define
dimensionless variables \begin{eqnarray} x&=&k({\eta_2} +{\eta_{\rm
e}}+\lambda),\nonumber\\ b&=&k({\eta_{\rm obs}}+{\eta_2} ).
\label{xbdefinition} \end{eqnarray} In terms of these variables, the
multipole moments (\ref{aldefinition}) are given by \begin{equation}
\langle a_l^2\rangle=24 \pi^3{\rho_{\rm ds}\over \rho_{\rm
p}}(l-1)l(l+1)(l+2)\int_0^\infty {dy\over y} \tilde{I}_l^2(y),
\label{alstarobinsky} \end{equation} where \begin{equation}
\tilde{I}_l(y)=\int_{\varepsilon y}^y dx {J_{5/2}(x)\over
x^{3/2}}{J_{l+1/2}(y-x)\over (y-x)^{5/2}}, \label{ilstarobinsky}
\end{equation} and \begin{equation} \varepsilon={{\eta_2} +{\eta_{\rm
e}}\over {\eta_{\rm obs}}+{\eta_2} }=(1+Z_{\rm ls})^{-1/2}.
\label{omegadefinition} \end{equation} Equations (\ref{alstarobinsky})
and (\ref{ilstarobinsky}) are equivalent to equation (8) in reference
\cite{Starobinsky2}.  The lower limit of the integral in
(\ref{ilstarobinsky}) appears different since the conformal time in our
scale factor (\ref{scalefactor}) during
the matter phase is shifted from that
in \cite{Starobinsky2} by the constant ${\eta_2} $.


\section{progress towards a closed form \protect\\
for the angular correlation function $C(\gamma)$}
\label{section5}

Equation (\ref{closedformscalar}) is a closed form for the angular
correlation function due to {\it scalar} perturbations. In this
section, we attempt to find a closed form for the angular correlation
function due to gravitational-wave perturbations.  Using
(\ref{ilk}),(\ref{aldefinition}), and (\ref{cofgamma}) one may write
the correlation function as \begin{equation}
C(\gamma)=\pi\int\limits_0^{{\eta_{\rm obs}}-{\eta_{\rm e}}}d\lambda
\int\limits_0^{{\eta_{\rm obs}}-{\eta_{\rm e}}}d\lambda'\int_0^\infty
{dk\over k} {F(\lambda,k) F^*(\lambda',k)\over k^2 D^2(\lambda)
D^2(\lambda')} B(D(\lambda),D(\lambda'),k,\gamma),
\label{correlationfunction} \end{equation} where \begin{equation}
B(r,r',k,\gamma)\equiv \sum_{l=0}^\infty (2l+1){(l+2)!\over (l-2)!}
j_l(k r)j_l(k r') P_l(\cos\gamma), \label{Bdefinition} \end{equation}
and $D(\lambda)$ is  defined in (\ref{Ddefinition}). To obtain a closed
form for the correlation function one must complete the  integrals over
$\lambda$, $\lambda'$, and $k$, and the infinite sum over $l$.

\subsection{The Sum Over $l$}

One can sum over $l$ and find a closed form for $B(r,r',k,\gamma)$
using an addition theorem for spherical Bessel functions. Consider the
addition theorem (see equation 10.1.45 in \cite{AbramowitzStegun})
\begin{equation} {\sin ks\over ks}=\sum_{l=0}^\infty (2l+1) j_l(kr)
j_l(kr') P_l(\cos\gamma), \label{additiontheorem} \end{equation} where
the length $s$ is defined by the non-negative root of \begin{equation}
s^2=r^2+r'^2-2 r r'\cos\gamma.  \end{equation} The right-hand side of
(\ref{additiontheorem}) is the same as the right-hand side of
(\ref{Bdefinition}), apart from the ratio of factorials. The ratio of
factorials is just a fourth order polynomial in $l$. To generate this
polynomial we define the derivative operator \begin{equation} {\cal
P}\equiv{1\over\sin\gamma}{\partial\over\partial\gamma}
\sin\gamma{\partial\over\partial\gamma}, \end{equation} which is the
Laplacian on the unit two-sphere for functions with azimuthal
symmetry.  The Legendre polynomials are eigenfunctions of ${\cal P}$,
and obey \begin{equation} {\cal
P}P_l(\cos\gamma)=-l(l+1)P_l(\cos\gamma).  \end{equation} Using this
one can quickly show that \begin{equation} {\cal P}({\cal P}+2)
P_l(\cos\gamma)= {(l+2)!\over (l-2)!}P_l(\cos\gamma).  \label{PonPl}
\end{equation} Using the addition theorem (\ref{additiontheorem}) and
(\ref{PonPl}) one obtains \begin{equation} B(r,r',k,\gamma)={\cal
P}({\cal P}+2){\sin ks\over ks}.  \label{Bresult} \end{equation} One
may distribute the derivative operator ${\cal P}({\cal P}+2)$ on $\sin
ks/ks$ to obtain a closed form for $B(r,r',k,\gamma)$. We prefer not to
distribute the derivative operator, and instead use
(\ref{correlationfunction}) and (\ref{Bresult}) to write the
correlation function as \begin{equation}
C(\gamma)=\pi\int\limits_0^{{\eta_{\rm obs}}-{\eta_{\rm e}}}d\lambda
\int\limits_0^{{\eta_{\rm obs}}-{\eta_{\rm e}}}d\lambda'\int_0^\infty
{dk\over k} {F(\lambda,k) F^*(\lambda',k)\over k^2 D^2(\lambda)
D^2(\lambda')} {\cal P}({\cal P}+2){\sin
ks(\lambda,\lambda',\gamma)\over ks(\lambda,\lambda',\gamma)},
\label{correlationfunction2} \end{equation} where now \begin{equation}
s(\lambda,\lambda',\gamma)\equiv \sqrt{D^2(\lambda)+D^2(\lambda') -2
D(\lambda) D(\lambda')\cos\gamma}.  \label{sdefinition2} \end{equation}
This form of the correlation function is very general. It only depends
upon the cosmological model through the graviton mode function (or more
precisely, its first derivative) which appears as $F(\lambda,k)$.

\subsection{The Integral Over $k$}

The next step to finding a closed form expression for the correlation
function is to evaluate the integral over the wavenumber $k$.  Since
the derivative of the graviton mode function depends on the wavenumber
$k$, one can not integrate over $k$ without using a specific form for
$F(\lambda,k)$.  We use the long wavelength approximation
(\ref{Fapproximation}), and assume it valid for all wavenumbers $k$.
The accuracy of this assumption is discussed in section \ref{section6};
the principle conclusion is that $C(\gamma)$ will be accurate for
$\gamma$ greater than a few degrees.  Substituting the long wavelength
approximation into (\ref{correlationfunction2}) one obtains
\begin{eqnarray} C(\gamma)=&&12{\rho_{\rm ds}\over\rho_{\rm p}}
\int\limits_0^{{\eta_{\rm obs}}-{\eta_{\rm e}}}d\lambda
\int\limits_0^{{\eta_{\rm obs}}-{\eta_{\rm e}}}d\lambda'
\{D^2(\lambda)D^2(\lambda')R(\lambda)R(\lambda')\}^{-1}\nonumber\\
&&\times \int_0^\infty {dk\over k} {j_2(kR(\lambda))
j_2(kR(\lambda'))\over k^4}{\cal P}({\cal P}+2) {\sin
ks(\lambda,\lambda',\gamma)\over ks(\lambda,\lambda',\gamma)},
\end{eqnarray} where \begin{equation} R(\lambda)\equiv{\eta_2}
+{\eta_{\rm e}}+\lambda.  \end{equation} Note that the only dependence
of the right-hand side on $\gamma$ is in the derivative operator ${\cal
P}({\cal P}+2)$, and in $s(\lambda,\lambda',\gamma)$.

The derivative operator ${\cal P}({\cal P}+2)$ is independent of $k$,
so one might wish to take it outside the $k$-integral.  The remaining
integrand could then be recast as a sum of trigonometric functions
times powers of $k$. The problem with this is that the resulting
integral over $k$ is logarithmically divergent because the remaining
integrand diverges as $k^{-1}$ for small $k$.

Still, one may take the derivative operator outside the integral by
setting the lower limit to a small, positive number $\epsilon$.  After
applying the operator ${\cal P}({\cal P}+2)$ one can then take the
limit as $\epsilon$ vanishes.  So one can write the correlation
function as
\begin{eqnarray}
C(\gamma)=&&12{\rho_{\rm ds}\over\rho_{\rm p}}
\int\limits_0^{{\eta_{\rm obs}}-{\eta_{\rm e}}}d\lambda
\int\limits_0^{{\eta_{\rm obs}}-{\eta_{\rm e}}}d\lambda'
\{D^2(\lambda)D^2(\lambda') R(\lambda)R(\lambda')\}^{-1}
\nonumber\\
&&\times\lim_{\epsilon\rightarrow 0}\>
{\cal P}({\cal P}+2)K_\epsilon(R(\lambda),R(\lambda'),
s(\lambda,\lambda',\gamma)),
\label{correlationfunction4}
\end{eqnarray}
where
\begin{equation}
K_\epsilon(a,b,c(\gamma))\equiv\int_\epsilon^\infty dk
\>{j_2(ka)j_2(kb)\over k^5}{\sin kc(\gamma)\over kc(\gamma)}.
\end{equation}
The function $K_\epsilon$ is well defined and finite for $\epsilon>0$;
one may evaluate it using standard techniques.

To evaluate $K_\epsilon$, express the spherical Bessel functions as
exponential functions divided by powers \cite{Jackson}, expand the
integrand, and integrate term by term (see 2.324.2 of reference
\cite{GradshteynRyzhik}).  This yields
\begin{equation}
K_\epsilon(a,b,c(\gamma))=-{a^2 b^2\over 225}\ln\epsilon
+U(a,b)+V(a,b,c(\gamma))+{\rm O}(\epsilon),
\label{KwithUV}
\end{equation}
where the functions $U$ and $V$ are independent of $\epsilon$, and
terms which vanish as $\epsilon$ goes to zero are not explicitly shown.
The term proportional to $\ln\epsilon$ and $U(a,b)$ do not depend on
$\gamma$ and are annhilated by ${\cal P}({\cal P}+2)$. Only the
function $V(a,b,c(\gamma))$ contributes to the correlation function
$C(\gamma)$. The function $V(a,b,c(\gamma))$ is a sum of more
than 25 terms,
each of which is a rational function of $a,b,$ and $c$, or a
rational function of $a,b,$ and $c$ times $\ln|p(a,b,c)|$, where
$p(a,b,c)$ is a second order polynomial in $a,b$, and $c$.
Using (\ref{correlationfunction4}) and (\ref{KwithUV}) one
can write the angular correlation function as
\begin{equation}
C(\gamma)=12{\rho_{\rm ds}\over\rho_{\rm p}}
\int\limits_0^{{\eta_{\rm obs}}-{\eta_{\rm e}}}d\lambda
\int\limits_0^{{\eta_{\rm obs}}-{\eta_{\rm e}}}d\lambda'\>
{\cal P}({\cal P}+2){V(R(\lambda),R(\lambda'),
s(\lambda,\lambda',\gamma))\over
D^2(\lambda)D^2(\lambda') R(\lambda)R(\lambda')},
\label{CwithV}
\end{equation}
where we have neglected to explicitly write out the
function $V$, since it is not very illuminating.

\subsection{The Integrals Over $\lambda$ and $\lambda'$}

The final step in finding a closed form for the angular correlation
function is to complete the remaining integrals over $\lambda$ and
$\lambda'$.  These integrals, however, are difficult for a number of
reasons. Distributing the derivative operator ${\cal P}({\cal P}+2)$
over the integrand of (\ref{CwithV}) yields on the order of
1000 terms. A large number of these terms are proportional to a
logarithm, with argument linear in the function $s(\lambda,\lambda')$.
The function $s(\lambda,\lambda')$ is the square root of a second-order
polynomial in $\lambda$ and $\lambda'$ which is not  factorable for
arbitrary $\gamma$.  Integrating terms like these over $\lambda$ and
$\lambda'$ is not trivial. Other terms are proportional to odd powers
of $s(\lambda,\lambda')$, and are  difficult for the same reason.  The
total number of terms, combined with the difficulty of integrating each
term, impedes further progress.

Other methods for finding a closed form for the angular correlation
function do not appear more promising. One can write $K_\epsilon$ in
the limit as $\epsilon$ vanishes as a hypergeometric function of
{\it two} variables (see 6.578.1 in reference \cite{GradshteynRyzhik}),
but again the remaining integrals over $\lambda$ and $\lambda'$ are
difficult. The integral over the wavenumber $k$ can be evaluated before
summing over $l$, though this involves the integral of four Bessel
functions, each with a different argument. One may also consider the
integrals over $\lambda$ and $\lambda'$ first. These integrals are
almost, but not quite, standard integral transforms of Bessel
functions. Another approach is to begin with the Sachs-Wolfe operator
(\ref{qsw}) and calculate the angular correlation function
(\ref{qcorr}) directly without any expansions in terms of spherical
harmonics.  This approach, however, reproduces
(\ref{correlationfunction2}).


\section{comparison of the exact and long wavelength\protect\\
approximate multipole moments $\langle
\lowercase{a}_{\lowercase{l}}^2
\rangle$}
\label{section6}

\subsection{Analytical Comparison}

We are considering an inflationary cosmological model that begins with
a de Sitter phase followed by radiation- and then  matter-dominated
phases. The graviton mode function (\ref{finalmodefunction}) that we
obtained for this model is {\it exact}, and valid for all  wavenumbers
$k$.  Using this mode function  one can calculate the multipole moments
$\langle a_l^2\rangle$ an observer in this universe model would
measure.

In the standard literature
\cite{Rubakov,Fabbri,AbbotWise,Starobinsky2,AbbottHarari,Turner,KraussWhite,White,White2}
, however, the long wavelength
{\it approximate} mode function (\ref{phiapprox}), rather than the
exact mode function (\ref{finalmodefunction}), is used to calculate the
multipole moments. The approximate mode function is only valid for
longer wavelengths, so we expect the angular correlation function, when
calculated using the approximate mode function, to be accurate only on
large angular scales.  Equivalently, we expect the moments $\langle
a_l^2\rangle$ calculated using the long wavelength approximate mode
function to only be accurate for small $l$.

The  straightforward way to determine for which $\langle a_l^2\rangle$
the long wavelength approximation is valid is to numerically
calculate    the moments using both the exact and approximate mode
functions, and compare. To calculate the moments one substitutes into
(\ref{ilk}) either the derivative (\ref{Fapproximation}) of the
approximate mode function, or the derivative of the exact mode function
(\ref{finalmodefunction}). One then uses (\ref{aldefinition}) to obtain
the multipole moments. Making these substitutions, one finds for the
moments calculated with the long wavelength approximate mode function
\begin{equation} \langle a_l^2\rangle_{\text{long wavelength
approximation}}=48\pi^2{(l+2)!\over (l-2)!}{\rho_{\rm ds}\over\rho_{\rm
p}}\int_0^\infty dy\>y^3 {\tilde J}_l^2(y), \label{alapprox}
\end{equation} and for the moments calculated with the exact mode
function \begin{equation} \langle a_l^2\rangle=48\pi^2{(l+2)!\over
(l-2)!}{\rho_{\rm ds}\over\rho_{\rm p}}\int_0^\infty
dy\>y^3\{\Upsilon_1(y){\tilde J}_l^2(y)+\Upsilon_2(y) {\tilde
Y}_l^2(y)-\Upsilon_3(y){\tilde J}_l(y){\tilde Y}_l (y)\}.
\label{alexact} \end{equation} The dimensionless variables $x$ and $y$
are defined by the change of variables \begin{equation}
x\equiv{{\eta_{\rm obs}}-{\eta_{\rm e}}-\lambda\over{\eta_{\rm
obs}}-{\eta_{\rm e}}} \>\>\text{and}\>\> y\equiv k({\eta_{\rm
obs}}-{\eta_{\rm e}}).  \label{ydefinition} \end{equation} The
functions ${\tilde J}_l(y)$ and ${\tilde Y}_l(y)$ are defined by
\begin{equation} {\tilde J}_l(y) \equiv\int_0^1 dx {J_{l+1/2}(yx)\over
(yx)^{5/2}} {j_2(y(\xi-x))\over y^2(\xi-x)} \>\>\text{and}\>\> {\tilde
Y}_l(y) \equiv\int_0^1 dx {J_{l+1/2}(yx)\over (yx)^{5/2}}
{y_2(y(\xi-x))\over y^2(\xi-x)}, \label{scriptjy} \end{equation} where
$\xi$ is a dimensionless constant determined by the redshift of the
last scattering surface $Z_{\rm ls}$:  \begin{equation} \xi={{\eta_{\rm
obs}}+{\eta_2} \over{\eta_{\rm obs}}-{\eta_{\rm e}}}=[1-(1+Z_{\rm
ls})^{-1/2}]^{-1}.  \label{xidefinition} \end{equation} In realistic
cosmological models, $\xi$ is slightly greater than one.  The three
functions $\Upsilon_i(y)$ depend on the Bogolubov coefficients $\alpha$
and $\beta$, and are defined as \begin{eqnarray}
\Upsilon_1(y)&\equiv&\bigg|{4\over 3}y^3\zeta_1^2\zeta_2(\alpha +\beta
)\bigg|^2,\\ &&\nonumber\\ \Upsilon_2(y)&\equiv&\bigg|{4\over
3}y^3\zeta_1^2\zeta_2(\alpha -\beta )\bigg|^2,\\ &&\nonumber\\
\Upsilon_3(y)&\equiv&{64\over 9}y^6\zeta_1^4\zeta_2^2{\rm Im}\{\alpha^*
\beta \}.  \label{upsilon3definition} \end{eqnarray} The Bogolubov
coefficients are given by (\ref{alphabetadefinition}) with the changes
of variable (\ref{ydefinition}).  The dimensionless constants $\zeta_1$
and $\zeta_2$ are determined by the redshifts $Z_{\rm end}$, $Z_{\rm
equal}$, and $Z_{\rm ls}$ defined in
(\ref{Zenddefinition}-\ref{Zlsdefinition}):  \begin{eqnarray}
\zeta_1&\equiv&{{\eta_1}\over{\eta_{\rm obs}}-{\eta_{\rm e}}}= {1\over
2}\bigg({\sqrt{1+Z_{\rm ls}}\over \sqrt{1+Z_{\rm
ls}}-1}\bigg){\sqrt{1+Z_{\rm equal}}\over (1+Z_{\rm end})},\\
&&\nonumber\\ \zeta_2&\equiv&{{\eta_2} \over{\eta_{\rm obs}}-{\eta_{\rm
e}}}= {1\over 2} \bigg({\sqrt{1+Z_{\rm ls}}\over \sqrt{1+Z_{\rm
ls}}-1}\bigg) {1\over \sqrt{1+Z_{\rm equal}}}.  \label{zeta2definition}
\end{eqnarray} To determine for which $\langle a_l^2\rangle$ the long
wavelength approximation is valid one numerically integrates
(\ref{alapprox}) and (\ref{alexact}) and compares the values for the
moments.

For cosmological models with ``enough'' inflation to solve the horizon
and flatness problems, $\zeta_1$ is very small since $Z_{\rm
end}>10^{26}$. For this reason one can approximate the $\Upsilon_i(y)$
by \begin{eqnarray} \Upsilon_1(y)&=&{\big(4 y\zeta_2\cos
(y\zeta_2)-\sin (y\zeta_2) +8 y^2 \zeta_2^2\sin (y\zeta_2)+\sin (3
y\zeta_2)\big)^2\over 36 y^2\zeta_2^2}+{\rm
O}(\zeta_1),\label{upsilon11}\\ &&\nonumber\\ \Upsilon_2(y)&=&{\big(-4
y\zeta_2\sin (y\zeta_2)-\cos (y\zeta_2) +8 y^2 \zeta_2^2\cos
(y\zeta_2)+\cos (3 y\zeta_2)\big)^2\over 36 y^2\zeta_2^2}+{\rm
O}(\zeta_1),\label{upsilon21}\\ &&\nonumber\\ \Upsilon_3(y)&=&{1\over 9
y^2\zeta_2^2}\bigg\{-\sin^2 (y\zeta_2)\sin (4 y\zeta_2) -4 y\zeta_2
(1+2\cos(2 y\zeta_2))\sin^2 (y\zeta_2) -32 y^2\zeta_2^2\cos
(y\zeta_2)\nonumber\\ &&+16 y^3\zeta_2^3\cos (2 y\zeta_2) +16
y^4\zeta_2^4\sin (2 y\zeta_2)\bigg\}+{\rm O}(\zeta_1).
\label{upsilon31} \end{eqnarray} In what follows, we neglect the ${\rm
O}(\zeta_1)$ and higher terms in $\Upsilon_i(y)$.  Note that the
standard long wavelength approximation (\ref{alapprox}) is equivalent
to setting $\Upsilon_1(y)=1$ and $\Upsilon_2(y)=\Upsilon_3(y)=0$ in the
exact expression (\ref{alexact}).  Indeed, expanding (\ref{upsilon11}),
(\ref{upsilon21}), and (\ref{upsilon31}) as power series in $y$ one
finds \begin{eqnarray} \Upsilon_1(y)&=&1+{\rm
O}(y\zeta_2)^2,\label{upsilon12}\\ &&\nonumber\\
\Upsilon_2(y)&=&{256\over 18225} (y\zeta_2)^{10}+{\rm
O}(y\zeta_2)^{12},\label{upsilon22}\\ &&\nonumber\\
\Upsilon_3(y)&=&{32\over 135}(y\zeta_2)^5+{\rm O}(y\zeta_2)^7.
\label{upsilon32} \end{eqnarray} So for $y\zeta_2<1$, to a good
approximation one has $\Upsilon_1(y)=1$ and
$\Upsilon_2(y)=\Upsilon_3(y)=0$.

Using the power series (\ref{upsilon12}-\ref{upsilon32}) we can
understand why the standard approximation (\ref{alapprox}) {\it is} the
long wavelength approximation to (\ref{alexact}). The functions
${\tilde J}_l(y)$ and ${\tilde Y}_l(y)$ are peaked near $y=l$. Figure
\ref{figure1} shows ${\tilde J}_l(y)$ and ${\tilde Y}_l(y)$ for $l=10$
and $l=100$.  Hence, if $l<\zeta_2^{-1}$, ${\tilde J}_l(y)$ and
${\tilde Y}_l(y)$ only have support for $y<\zeta_2^{-1}$, which is the
same range for which $\Upsilon_1(y)\approx 1$ and $\Upsilon_2(y)
\approx \Upsilon_3(y)\approx 0$. For $y>\zeta_2^{-1}$, ${\tilde
J}_l(y)$ and ${\tilde Y}_l(y)$ have no support, and the second and
third terms in the integrand of the exact formula (\ref{alexact}) do
not contribute for large $y$. So one expects the standard approximation
(\ref{alapprox}) to give accurate values of $\langle a_l^2\rangle$ for
$l<\zeta_2^{-1}$. From (\ref{zeta2definition}) note that for realistic
models, $\zeta_2\approx Z_{\rm equal}^{-1/2}$.  So one expects that for
$l<Z_{\rm equal}^{1/2}$ the standard long wavelength approximation
gives accurate values for the multipole moments $\langle
a_l^2\rangle$.

\subsection{Numerical Comparison}

Table \ref{table2} lists the multipole moments for various $l$ values,
calculated using both the approximate formula (\ref{alapprox}) and the
exact formula (\ref{alexact}) for $Z_{\rm end}=10^{27}, Z_{\rm equal}=
10^4,$ and $Z_{\rm ls}=1300$.  The difference between the exact and
long wavelength approximate moments is shown in Figures \ref{figure2}-
\ref{figure4}
for different values of the cosmological parameters.
For $2\leq l\leq 10$ our values of $\langle a_l^2\rangle$ agree very
well with those of White \cite{White} (due to a difference in the
definition of $\langle a_l^2\rangle$, our results are smaller than
White's by a factor of $2l+1$).  As expected, for smaller $l$ the
values of $\langle a_l^2\rangle$ from the approximate formula are in
good agreement with the exact $\langle a_l^2\rangle$. For $l\leq 30$,
the difference between the exact and approximate moments is less than
two percent of the exact result. For $l\leq 100$, the difference is
less than twenty percent. When $l$ is 200, however, the disagreement is
more substantial; the exact value is more than twice the approximate
value. The disagreement is even more for larger $l$, and for $l=1000$,
the exact value is a factor of 69 larger than the approximate value.
The long wavelength approximate formula (\ref{alapprox}) substantially
{\it underestimates} the contribution of the large $l$ moments $\langle
a_l^2\rangle$ to the angular correlation function $C(\gamma)$.

Similar results, which reveal the shortcomings of the approximate
formula for $\langle a_l^2\rangle$ have been obtained by Turner, White
and Lidsey \cite{Turner2}.  Their approach is less analytical than our
own; they use numerical methods to solve the Klein-Gordon equation and
obtain exact mode functions $\phi(\eta,k)$ analogous to our equation
(\ref{finalmodefunction}).  They express these exact solutions in terms
of the standard long-wavelength approximate mode functions, using a
``transfer function".   Figure \ref{figure5} shows the results of our
best attempt to obtain the Turner, White, and Lidsey results from our
analytical formula, together with their published data.  By tuning the
parameters of our cosmological model to $Z_{\rm ls} = 900$ and $Z_{\rm
equal}=2500$, we have been able to obtain fairly close agreement
between the two sets of results.  One should note, however, that
Turner, White, and Lidsey consider a universe which is not completely
matter-dominated at the time when the CBR is emitted.  Their universe
model is more realistic than our own, since we have assumed the
universe to be completely matter-dominated at the time of last
scattering. Ng and Speliotopoulos \cite{Ng2}, using a WKB formalism,
have also considered a universe which transforms smoothly from the
radiation- to the matter-dominated phases.  Some of their results,
however, do not appear consistent with those of \cite{Turner2}.

We have shown that the long wavelength approximate formula
(\ref{alapprox}) substantially {\it underestimates} the contribution of
the large $l$ multipole moments $\langle a_l^2\rangle$ to the angular
correlation function $C(\gamma)$. Although this long wavelength
approximation has been used previously to interpret published
experimental data, one does not expect the new results presented here
to significantly affect the conclusions.  This is because for
reasonable values of the redshift $Z_{\rm equal}$, the discrepency
between the approximate  and exact results is significant only for
multipole moments  which one expects would be dominated by the
contribution from {\it scalar} perturbations \cite{Bond}. However,
Krauss and White \cite{KraussWhite} and Grishchuk \cite{Grishchuk3}
have suggested that the relative
contribution to the CBR anisotropy from
gravitational waves has been underestimated, and that these
contributions might dominate the multipole moments.


\section{conclusion}
\label{conclusion}

In this paper, we have shown how the rapid expansion of the universe
during an inflationary phase creates large numbers of gravitons, whose
collective effects produce potentially-observable fluctuations in the
temperature of the CBR.  The correlation function of these temperature
fluctuations may be calculated from first principles; for example the
overall magnitude of the perturbations is determined by the uncertainty
principle.  The exact expression that we obtain for the correlation
function agrees with standard published results for the lower multipole
moments, but has larger temperature fluctuations in the higher
multipole moments than  predicted by the standard published formulae.
This appears to be in good quantitative agreement with recently
published numerical work by Turner, White, and Lidsey\cite{Turner2}.
The larger predicted temperature fluctuations in the higher multipole
moments, however, most likely will not lead to a reinterpretation of
the experimentally observed data since it is generally expected that
the observed anisotropy for the higher multipole moments will be due
almost entirely to scalar, rather than tensor perturbations.

As mentioned in the introduction, the original discovery that a
rapidly-expanding universe could create relic gravitational waves was
made by Grishchuk \cite{GrishchukCreation}.  In recent work
\cite{Grishchuk1,Grishchuk2}, Grishchuk analyzed the temperature
fluctuations produced by these waves, using the techniques of quantum
optics.  In his analysis, the classical gravitational field
``interacts" with the gravitons and acts as a ``pumping" field.   This
leaves the gravitational field in a squeezed quantum state today.
Grishchuk stresses the importance of the resulting phase correlations
to the final form of $C(\gamma)$.

In our language, the (quantized) gravitational field is taken to be in
the vacuum state of the initial de Sitter phase.  (Note that we use the
``Heisenberg picture" of quantum fields in which the states do not
evolve with time, but the operators do; we also assume that a sucessful
inflationary stage leaves the universe indistinguishably close to the
de Sitter vacuum state).  Although we do not use any of the techniques
of non-linear quantum optics that Grishchuk advocates, we nevertheless
reproduce, as intermediate results, his final formulae for
$C(\gamma)$.  In particular, Grishchuk's formula (10) from reference
\cite{Grishchuk1} is the same as our equation
(\ref{correlationfunction2}) with $\gamma \to 0$.  Formula (11) from
reference \cite{Grishchuk1} is the same as our equations
(\ref{cexpan},\ref{cwitha}-\ref{psidefinition}),
and formulae (12-13) from reference \cite{Grishchuk1} is
the same as our equation
(\ref{correlationfunction2}) with the action of the operator ${\cal
P}({\cal P} + 2)$ expanded out.  We agree that the correlation between
phases {\it is} important; in the sense that for example in our
equation (\ref{upsilon3definition}) the value of $\Upsilon_3$ depends
upon the relative phase of the positive and negative frequency
wavefunctions.  However, we stress that results identical to
Grishchuk's may be obtained, as we have shown, using only the standard
machinery of linearized quantum fields in curved spacetime
\cite{ParkerCreation,BirrellDavies}.

\acknowledgments
We are grateful to Alexei Starobinsky, who suggested that one might be
able to obtain a closed form for $C(\gamma)$ as attempted in Section
\ref{section5}.  This work has been partially supported by NSF grant
PHY91-05935.


\appendix

\section{}
\label{appendixA}

In section \ref{section2subsection2}  we argued that based on the
isotropy of the initial state of the universe (which we took
to be the de Sitter vacuum state), and on the istropy of
the FRW model, one expects the angular correlation function to be
rotationally invariant. For this reason one may write the matrix
element ${\langle 0|{\bar C}_{pq}^\dagger \bar C_{lm}|0\rangle}$ as in
(\ref{shortmatrixelement}), and then use
(\ref{bigclm}) for ${\bar C}_{lm}$ to solve for $\langle a_l^2\rangle$.
In this appendix we sketch this calculation. A somewhat more
complicated version of this calculation may be found in \cite{White}.

The primary advantage of writing the matrix element as in
(\ref{shortmatrixelement}) is that it allows one to make a useful
choice of coordinates and evaluate the integrals over angular
variables. Using (\ref{bigclm}) for ${\bar C}_{lm}$ one obtains from
(\ref{shortmatrixelement}) \begin{eqnarray} \langle
a_l^2\rangle&=&{1\over 4}\int\limits^ {{\lambda_{\rm obs}}
}_{{\lambda_{\rm e}}  }d\lambda'\int\limits^{{\lambda_{\rm obs}}
}_{{\lambda_{\rm e}} }d\lambda \int {d^3k\over k}F(\lambda
',k')F^{*}(\lambda,k) \bigg[e_{ab}(\hat k{}^e) e^{*}_{cd}(\hat k^e)
+e^{*}_{ab}(\hat k{}^e) e_{cd}(\hat k^e) \bigg]\nonumber\\ & &\times
\int d\Omega_{\hat v}\int d\Omega_{ \hat u} Y_{lm} (\hat v^e)Y^*_{lm}
(\hat u^e)\hat v ^c\hat v^d\hat u^a\hat u^b
 e^{-i k^f(D(\lambda)\hat u_f-D(\lambda')\hat v_f)}, \label{appendixa1}
\end{eqnarray} where we have set $p=l$ and $q=m$ to eliminate the
Kronecker delta functions on the right-hand side of
(\ref{shortmatrixelement}).  Since by assumption both sides of the
equation above are independent of $m$, one may sum both sides  from
$m=-l$ to $m=l$.  Using the addition theorem for spherical harmonics
\cite{Jackson}, and cancelling factors of $(2l+1)$ on both sides, one
obtains \begin{eqnarray} \langle a_l^2\rangle&=&{1\over
16\pi}\int\limits^ {{\lambda_{\rm obs}} }_{{\lambda_{\rm e}}
}d\lambda'\int\limits^{{\lambda_{\rm obs}} }_{{\lambda_{\rm e}}
}d\lambda \int {d^3k\over k}F(\lambda ',k')F^{*}(\lambda,k)
\bigg[e_{ab}(\hat k{}^e) e^{*}_{cd}(\hat k^e) +e^{*}_{ab}(\hat k{}^e)
e_{cd}(\hat k^e) \bigg]\nonumber\\ & &\times \int d\Omega_{\hat v}\int
d\Omega_{ \hat u} P_l(\cos\gamma) \hat v ^c\hat v^d\hat u^a\hat u^b
 e^{-i k^f(D(\lambda)\hat u_f-D(\lambda')\hat v_f)}, \end{eqnarray}
where  the angle $\gamma$ is defined by \begin{equation}
\cos\gamma\equiv \hat u^a\hat v_a.  \end{equation} Note that we have
not yet made a specific choice of coordinates.

One is free to choose whatever coordinates one wants to compute the
integrals over the angular variables $\Omega_{\hat u},\Omega_{\hat v}$,
and $\Omega_{\hat k}$. In particular the choice of coordinates for
$\Omega_{\hat u}$ and $\Omega_{\hat v}$ may depend on the vector $k^c$.
We choose coordinates so that the vectors $\hat u^c$ and $\hat v^c$ are
written in terms of the $(\hat m^c,\hat n^c,\hat k^c)$ triad as
\begin{eqnarray} \hat u^c&=&\sin\theta_{\hat u}\cos\phi_{\hat u}\hat
m^a +\sin\theta_{\hat u}\sin\phi_{\hat u}\hat n^a +\cos\theta_{\hat
u}\hat k^a\\ \hat v^c&=&\sin\theta_{\hat v}\cos\phi_{\hat v}\hat m^a
+\sin\theta_{\hat v}\sin\phi_{\hat v}\hat n^a +\cos\theta_{\hat v}\hat
k^a.  \end{eqnarray} With this choice of coordinates, and using the
form of the polarization tensors given in (\ref{circpolar}), one can
quickly show that the contraction between the polarization tensors and
the unit vectors is \begin{equation} \bigg[e_{ab}(\hat k{}^e)
e^{*}_{cd}(\hat k^e) +e^{*}_{ab}(\hat k{}^e) e_{cd}(\hat k^e)
\bigg]\hat v ^c\hat v^d\hat u^a\hat u^b= \sin^2\theta_{\hat
u}\sin^2\theta_{\hat v}\cos(2 \phi_{\hat u}-2\phi_{\hat v}).
\label{contraction} \end{equation} Also using these coordinates one may
again use the addition theorem for spherical harmonics to write
\begin{equation} P_l(\cos\gamma)={4\pi\over (2l+1)}\sum_{m=-l}^l
Y^*_{lm}(\theta_{\hat u},\phi_{\hat u}) Y_{lm}(\theta_{\hat
v},\phi_{\hat v}), \label{additiontheorem2} \end{equation} where
\begin{equation} \cos\gamma=\cos\theta_{\hat u}\cos\theta_{\hat v}+
\sin\theta_{\hat u}\sin\theta_{\hat v}\cos(\phi_{\hat u}- \phi_{\hat
v}).  \end{equation} (Note the following subtle point. The
$Y^*_{lm}(\theta_{\hat u},\phi_{\hat u})$ and $Y_{lm}(\theta_{\hat
v},\phi_{\hat v})$ in the right-hand side of (\ref{additiontheorem2})
do not in general have the same values as the spherical harmonic
functions which appear in (\ref{appendixa1}) because we have done a
coordinate rotation that depends on $\hat k^c$.  In general these
values are related  by a linear expansion involving Clebsch-Gordon
coefficients.) Using (\ref{contraction}) and (\ref{additiontheorem2})
one finds for the multipole moment \begin{eqnarray} \langle
a_l^2\rangle&=&{1\over 16\pi}\int\limits^ {{\lambda_{\rm obs}}
}_{{\lambda_{\rm e}}  }d\lambda'\int\limits^{{\lambda_{\rm obs}}
}_{{\lambda_{\rm e}} }d\lambda \int {d^3k\over k}F(\lambda
',k')F^{*}(\lambda,k) \int d\Omega_{\hat v}\int d\Omega_{\hat u} e^{-i
k(D(\lambda)\cos\theta_{\hat u}-D(\lambda')\cos\theta_{\hat
v})}\nonumber\\ &&\times \sin^2\theta_{\hat u}\sin^2\theta_{\hat
v}\cos(2 \phi_{\hat u}-2\phi_{\hat v})\bigg\{{4\pi\over
(2l+1)}\sum_{m=-l}^l Y_{lm}^*(\theta_{\hat u},\phi_{\hat
u})Y_{lm}(\theta_{\hat v},\phi_{\hat v}) \bigg\}.  \end{eqnarray} In
this form one may evaluate all the integrals over angular variables.

The integrals over the angles $\theta_{\hat u},\theta_{\hat v},
\phi_{\hat u}$, and $\phi_{\hat v}$ can be done in a straightforward
way by writing the spherical harmonics as products of exponentials and
Legendre functions. With the definition \begin{equation}
Y_{lm}(\theta,\phi)\equiv\sqrt{{(2l+1)\over 4\pi}{(l-m)!\over
(l+m)!}}\>P_l^m(\cos\theta)e^{im\phi}, \end{equation} one finds after
integrating by parts $l-2$ times (formula 3.387.2 in Gradshteyn and
Ryzhik \cite{GradshteynRyzhik} is  helpful) \begin{eqnarray} \langle
a_l^2\rangle&=&{\pi\over 2}{(l+2)!\over (l-2)!}\int\limits^
{{\lambda_{\rm obs}} }_{{\lambda_{\rm e}}
}d\lambda'\int\limits^{{\lambda_{\rm obs}} }_{{\lambda_{\rm e}}
}d\lambda \int_0^\infty dk k\>F(\lambda ',k')F^{*}(\lambda,k)
{j_l(kD(\lambda))j_l(kD(\lambda'))\over k^4 D^2(\lambda)
D^2(\lambda')}\nonumber\\ &&\times\int_0^{2\pi} d\theta_{\hat
k}\sin\theta_{\hat k} \int_0^\pi d\phi_{\hat k}\>\sum_{m=-l}^l
(\delta_{m,2}+\delta_{m,-2}).  \end{eqnarray} The remaining integals
over $\theta_{\hat k}$ and $\phi_{\hat k}$ are trivial and yield
$4\pi$. The sum over $m$ is also trivial and contributes a factor of 2.
So one obtains for the multipole moment \begin{equation} \langle
a_l^2\rangle= 4\pi^2{(l+2)!\over (l-2)!} \int_0^\infty {dk\over k}
\int\limits^ {{\lambda_{\rm obs}} }_{{\lambda_{\rm e}}
}d\lambda'\int\limits^{{\lambda_{\rm obs}} }_{{\lambda_{\rm e}}
}d\lambda F(\lambda ',k')F^{*}(\lambda,k)
{j_l(kD(\lambda))j_l(kD(\lambda'))\over k^2 D^2(\lambda)
D^2(\lambda')}.  \end{equation} Recalling the definitions of
$D(\lambda)$ and $I_l(k)$, one can write this as \begin{equation}
\langle a_l^2\rangle= 4\pi^2{(l+2)!\over (l-2)!} \int_0^\infty {dk\over
k} |I_l(k)|^2.  \end{equation} This result is the same given in
(\ref{aldefinition}).

\section{}
\label{appendixB}

This appendix  describes  the numerical techniques
used in section \ref{section6}.

The primary numerical technique used to evaluate both the approximate
(\ref{alapprox}) and exact multipole moments (\ref{alexact}) is
numerical integration.  Both integrals over $y$ in (\ref{alapprox}) and
(\ref{alexact}) were done using a fifth order embedded
Runge-Kutta-Fehlberg algorithm with adaptive stepsize control
\cite{PressEtAl}.  Although formally the upper limit of the integral
extends to infinity, we only integrated until the remaining
contribution became negligible.  This is possible because the
integrands in  (\ref{alapprox}) and (\ref{alexact}) fall off at least
as fast as $y^{-2}$ for large $y$.  Special care must be taken in
determining when the remaining contribution is negligible since the
integrand {\it does} have periodic zeroes, even for large $y$.

Both  ${\tilde J}_l(y)$ and ${\tilde Y}_l(y)$ (or more precisely, these
functions multiplied by $y^{7/2}$) were also calculated using a fifth
order embedded Runge-Kutta-Fehlberg algorithm with adaptive stepsize
control. No special treatment is needed since both the upper and lower
limits are finite, and the integrands are well behaved.  The spherical
Bessel functions in the integrands of (\ref{scriptjy}) can be expressed
in terms of trig functions \cite{Jackson}, and evaluated
using standard machine routines.

The Bessel function with index $l+1/2$ was evaluated with the routine
``bessjy'' given in chapter 6 of
\cite{PressEtAl}.  Although this routine is very accurate and fairly
fast, we did not use it to calculate the value of the Bessel function
every time it was needed in the integration algorithms.  This is
because the argument of the Bessel function
is a function of both $x$ {\it and} $y$, and so
the argument  is unique for every step taken
while the integral over $x$ is being calculated; it is not possible to
store certain values and ``reuse'' them later. A typical integration to
find a single moment for a particular $l$ would easily require on the
order of $10^6$ calls to the routine bessjy.

To reduce the number of ``expensive'' calls to bessjy
we used a cubic spline interpolation scheme to
calculate the Bessel functions with index $l+1/2$. For each different
value of $l$, a table of Bessel functions evaluated at equally spaced
intervals $\Delta=1/32$ was tabulated using bessjy. This
interval is small enough so that  cubic spline interpolation gives
values accurate to at least one part in $10^6$.  The cubic spline was
done using the routines ``spline'' and ``splint'' from chapter 3 of
\cite{PressEtAl}.



\begin{figure}
\caption{The functions  $|{\tilde J}_l(y)|$ (solid curve)
and $|{\tilde Y}_l(y)|$ (dashed curve)
for $l=10$ and $100$, normalized
so that their maximum value is one.
For all curves  $Z_{\rm ls}=1300$.
Both ${\tilde J}_l(y)$ and ${\tilde Y}_l(y)$
peak fairly strongly near $y=l$,
which makes the long wavelength approximate formula
for the angular correlation funcion multipole moments
accurate for $l\protect\lesssim Z_{\rm equal}^{1/2}$.}
\label{figure1}
\end{figure}

\begin{figure}
\caption{Multipole moments $\langle a_l^2\rangle$ normalized to the
quadrupole moment $\langle a_2^2\rangle$, with
$M_l\equiv\bigg({\rho_{\rm p}\over\rho_{\rm ds}}\bigg)\bigg(
{l(l+1)\over 6}\bigg)\bigg({\langle a_l^2\rangle\over \langle
a_2^2\rangle} \bigg)$.  The upper curve is the correct result
calculated using the exact graviton mode function. The lower curve is
the result obtained from the standard formula found in the literature
using a long wavelength approximation to the graviton mode function.
Both curves have $Z_{\rm end}>10^{20}$, $Z_{\rm equal}=10^4$, and
$Z_{\rm ls}=1300$.} \label{figure2} \end{figure}

\begin{figure}
\caption{Multipole moments $\langle a_l^2\rangle$
normalized to the quadrupole moment $\langle a_2^2\rangle$
with $M_l$ the same as in Figure \protect\ref{figure2}.
All three curves are  calculated using
the exact graviton mode function, and  have
$Z_{\rm end}>10^{20}$ and $Z_{\rm equal}=10^4$. The upper
curve has $Z_{\rm ls}=1300$, the middle curve $Z_{\rm ls}=800$,
and the lower curve $Z_{\rm ls}=400$.}
\label{figure3}
\end{figure}

\begin{figure}
\caption{Ratio of  multipole
moments obtained with the long wavelength approximation
to the exact multipole moments with $R_l\equiv
{\langle a_l^2\rangle}_{\rm long\>\, wavelength\>\, approx}/
\langle a_l^2\rangle$. All three curves have the same redshifts
as in Figure \protect\ref{figure3}. The approximate moments fail
to be accurate for $l>Z_{\rm equal}^{1/2}$.}
\label{figure4}
\end{figure}

\begin{figure}
\caption{Multipole coefficients $\langle a_l^2\rangle$ normalized to
the quadrupole moment $\langle a_2^2\rangle$, with $M_l$ the same as in
Figure \protect\ref{figure2}.  The diamonds show the results of Turner,
White, and Lidsey \protect\cite{Turner2} obtained by expressing exact
mode functions (obtained by numerically integrating the massless
Klein-Gordon equation)  in terms of the standard long-wavelength
approximate mode functions using a ``transfer function".  The upper
curve shows an exact result obtained from our analytic formula
(\protect\ref{alexact}) with $Z_{\rm ls}=900$ and $Z_{\rm equal}=2500$.
These parameters were chosen because they appeared to give the best
match to the Turner, White, and Lidsey result. The lower curve is the
result obtained from the standard long wavelength formula
(\protect\ref{alapprox}) with the same parameters used for the upper
curve.}
\label{figure5}
\end{figure}


\begin{table}
\caption{Change of dependent and independent
variable needed to cast the massless Klein-Gordon
equation in the form of Bessel's
differential equation, and
positive frequency (unnormalized) solution.}
\begin{tabular}{cccc}
Epoch&Dependent&Independent&Solution $\phi$\\
\tableline
&&&\\
$-\infty<\eta<{\eta_1}$& $\phi=(\eta-2{\eta_1})^2\chi$&
$z=k(\eta-2{\eta_1})$&$(\eta-2{\eta_1})^2
h_1^{(2)}(k(\eta-2{\eta_1}))$\\
&&&\\
${\eta_1}<\eta<{\eta_2} $& $\phi=\chi$&$z=k\eta$&
$h_0^{(2)}(k\eta)$\\
&&&\\
${\eta_2} <\eta$&$\phi=(\eta+{\eta_2} )^{-1}\chi$&
$z=k(\eta+{\eta_2} )$&
$(\eta+{\eta_2} )^{-1}h_1^{(2)}(k(\eta+{\eta_2} ))$\\
\end{tabular}
\label{table1}
\end{table}

\begin{table}
\caption{Multipole coefficients $\langle a_l^2\rangle$
for various $l$ predicted for a stochastic background
of gravitational radiation generated by exponential inflation.
Exact values are calculated using the exact graviton mode
function in  (\protect{\ref{aldefinition}}) for the multipole
moments. Approximate values are calculated using the standard long
wavelength approximation to the graviton mode function.
The values in this table are for redshifts $Z_{\rm end}>
10^{20}$, $Z_{\rm eq}=10^4$, and $Z_{\rm ls}=1300$.}
\begin{tabular}{cdd}
$l$ & Exact & Approximate\\
\tableline
&&\\
2&1.55&1.55\\
3&6.07$\times 10^{-1}$&6.07$\times 10^{-1}$\\
4&3.44$\times 10^{-1}$&3.44$\times 10^{-1}$\\
5&2.27$\times 10^{-1}$&2.27$\times 10^{-1}$\\
6&1.62$\times 10^{-1}$&1.62$\times 10^{-1}$\\
7&1.23$\times 10^{-1}$&1.22$\times 10^{-1}$\\
8&9.61$\times 10^{-2}$&9.59$\times 10^{-2}$\\
9&7.75$\times 10^{-2}$&7.73$\times 10^{-2}$\\
10&6.38$\times 10^{-2}$&6.36$\times 10^{-2}$\\
25&1.08$\times 10^{-2}$&1.07$\times 10^{-2}$\\
50&2.11$\times 10^{-3}$&2.01$\times 10^{-3}$\\
75&5.39$\times 10^{-4}$&4.85$\times 10^{-4}$\\
100&1.22$\times 10^{-4}$&1.01$\times 10^{-4}$\\
250&6.45$\times 10^{-7}$&1.47$\times 10^{-7}$\\
500&5.98$\times 10^{-8}$&3.49$\times 10^{-9}$\\
750&1.49$\times 10^{-8}$&4.93$\times 10^{-10}$\\
1000&5.07$\times 10^{-9}$&7.33$\times 10^{-11}$\\
\end{tabular}
\label{table2}
\end{table}

\end{document}